\documentclass[%
 reprint,
 amsmath,amssymb,
 aps,
 pre,
]{revtex4-2}
\usepackage[T1]{fontenc}
\usepackage[latin9]{inputenc}
\usepackage{graphicx}
\usepackage{esint}
\usepackage{xcolor}
\usepackage{bm}
\usepackage{babel}
\makeatletter

\providecommand{\tabularnewline}{\\}

\makeatother

\begin{document}
\title{Zero temperature phase transitions and their anomalous influence on thermodynamic behavior in the $q$-state Potts model on a diamond chain}
\author{Yury Panov}
\affiliation{Department of Theoretical and Mathematical Physics, Institute of Natural Sciences and Mathematics, Ural Federal University, 19 Mira street, 620002 Ekaterinburg, Russia}
\author{Onofre Rojas}
\affiliation{Department of Physics, Institute of Natural Science, Federal University
of Lavras, 37200-900 Lavras-MG, Brazil}
\begin{abstract}
The $q$-state Potts model on a diamond chain has mathematical significance in analyzing phase transitions and critical behaviors in diverse fields, including statistical physics, condensed matter physics, and materials science. By focusing on the 3-state Potts model on a diamond chain, we reveal rich and analytically solvable behaviors without phase transitions at finite temperatures. Upon investigating thermodynamic properties such as internal energy, entropy, specific heat, and correlation length, we observe sharp changes near zero temperature. Magnetic properties, including magnetization and magnetic susceptibility, display distinct behaviors that provide insights into spin configurations in different phases. However, the Potts model lacks genuine phase transitions at finite temperatures, in line with the Peierls argument for one-dimensional systems. Nonetheless, in the general case of an arbitrary $q$-state, magnetic properties such as correlation length, magnetization, and magnetic susceptibility exhibit intriguing remnants of a zero-temperature phase transition at finite temperatures. Furthermore, residual entropy uncovers unusual frustrated regions at zero-temperature phase transitions. This feature leads to the peculiar thermodynamic properties of phase boundaries, including a sharp entropy change resembling a first-order discontinuity without an entropy jump, and pronounced peaks in second-order derivatives of free energy, suggestive of a second-order phase transition divergence, but without singularities. This unusual behavior is also observed in the correlation length at the pseudo-critical temperature, which could potentially be misleading as a divergence.
\end{abstract}
\maketitle

\section{Introduction}

The one-dimensional Potts model, while simpler than higher-dimensional
models, exhibits a range of intriguing properties making it a focus
of study. It can often be solved exactly, offering valuable insight
into statistical systems without the need for approximations or numerical
methods \cite{wu1982potts}. These models lay the groundwork for understanding
more complex behaviors in higher dimensions and are central to the
study of phenomena like phase transitions in statistical physics \cite{baxter2016exactly}.
They can represent a variety of physical and mathematical systems,
such as counting colored planar maps problems \cite{bernardi}. Additionally,
they provide a practical platform for testing new computational methods,
including Monte Carlo algorithms and machine learning techniques applied
to statistical physics \cite{binder1981monte}.

Even though a finite-temperature phase transition is absent in one-dimensional
models with short range interaction, it is still feasible to define and study a \textquotedbl pseudo-critical\textquotedbl{}
temperature. This is commonly perceived as the temperature at which
a system fluctuations reach a peak, often associated with the system
specific heat, which generally exhibits a peak at the pseudo-critical
temperature. In this sense, recent research has unveiled a series
of decorated one-dimensional models, notably the Ising and Heisenberg
models, each exhibiting a range of structures. Among these are the
Ising-Heisenberg diamond chain \cite{torrico,torrico2}, the one-dimensional
double-tetrahedral model with a nodal site comprising a localized
Ising spin alternating with a pair of mobile electrons delocalized
within a triangular plaquette \cite{Galisova}, the ladder model with
an Ising-Heisenberg coupling in alternation \cite{on-strk}, and the
triangular tube model with Ising-Heisenberg coupling \cite{strk-cav}.
Pseudo-transition phenomena were detected in all these models. While
the first derivative of the free energy, like entropy, internal energy,
or magnetization demonstrates a jump akin to an abrupt change when
the temperature varies, the function remains continuous. This pattern
mimics a first-order phase transition. Nevertheless, a second-order
derivative of free energy, such as the specific heat and magnetic
susceptibility, showcases behavior typical of a second-order phase
transition at a finite temperature. This peculiar behavior has drawn
focus for a more meticulous study, as discussed in reference \cite{pseudo}.
More recently, reference \cite{Isaac} has provided additional dialogue
on this property and an exhaustive study of the correlation function
for arbitrarily distant spins surrounding the pseudo-transition. Furthermore,
certain conditions were proposed to observe the pseudo-transition,
which is associated with residual entropy \cite{Rojas2020-1,Rojas2020-2}. 

Recent discoveries have positioned azurite {[}$\mathrm{Cu_{3}(CO_{3})_{2}(OH)_{2}}${]}
as an intriguing quantum antiferromagnetic model, as described by
the Heisenberg model on a diamond chain. This has led to numerous
riveting theoretical investigations into diamond chain models. Notably,
Honecker et al. \cite{honecker} probed the dynamic and thermodynamic
traits of this model, while comprehensive analysis was conducted on
the thermodynamic attributes of the Ising-Heisenberg model on diamond-like
chains \cite{canova06,lisnii-11,vadim,valverde-1,valverde-2}. Additional studies
into the Ising-XYZ diamond chain model were inspired by current research,
including experimental explorations of the natural mineral azurite
and theoretical calculations of the Ising-XXZ model. Particular attention
was drawn by the appearance of a 1/3 magnetization plateau and a double
peak in both magnetic susceptibility and specific heat in experimental
measurements \cite{rule,kikuchi-1,kikuchi-2}. It is relevant to note that the
dimer interactions (interstitial sites) exhibit considerably stronger
exchange interaction than the nodal sites in $xy$-axes, especially
in the $z$-component. Consequently, this model can be accurately
represented as an exactly solvable Ising-Heisenberg model. Further
supporting this, experimental data regarding the magnetization plateau
align with the approximated Ising-Heisenberg model \cite{canova06,ananikian,chakh}.

In the context of one-dimensional Potts models, Sarkanych et al. \cite{sarkanych}
introduced a variation featuring invisible states and short-range
coupling. The notion of \textquotedbl invisible\textquotedbl{} in
this context refers to an additional level of energy degeneracy that
contributes solely to entropy without affecting interaction energy,
thus catalyzing the first-order phase transition. This proposal was
inspired by low-dimensional systems such as the simple zipper model~\cite{zimm-bragg}, 
a descriptor of long-chain DNA nucleotides. To 
account for narrow helix-coil transitions within these systems, Zimm  
and Bragg~\cite{zimm-bragg} put forth a largely phenomenological 
cooperative parameter. This innovative approach has since sparked 
numerous inquiries \cite{Badasyan10,ananikyan,Badasyan13,tonoyan}. 
In one-dimensional cooperative systems, Potts-like models \cite{Badasyan10,Badasyan13} 
serve as an effective representation, providing study of helix-coil 
transitions in polypeptides~\cite{ananikyan} --- a classic application 
of theoretical physics to macromolecular systems, yielding insightful 
comprehension of helix-coil transition properties. The reversible 
adsorption demonstrated by polycyclic aromatic surface elements in 
carbon nanotubes (CNTs) and aromatic DNA further enriches these studies. 
To consider DNA-CNT interactions, Tonoyan et al.~\cite{tonoyan} adjusted 
the Hamiltonian of the zipper model~\cite{zimm-bragg}. Similarly, 
our earlier work~\cite{Panov} proposed a one-dimensional Potts model 
combined with the Zimm-Bragg model, what we call here simply as Potts-Zimm-Bragg 
model, as a result leading to the observation of several distinctive 
properties. 

The paper is structured as follows: Section 2 presents our proposal
for a $q$-state Potts model on a diamond chain structure. Section
3 analyzes the zero-temperature phase transition, residual entropy,
and corresponding magnetizations. Section 4 discusses the thermodynamic
solution for finite $q$-states and explores physical quantities such
as entropy, magnetization, specific heat, magnetic susceptibility,
and correlation length. This section also highlights the presence
of pseudo-critical temperatures. Finally, Section 5 summarizes our
findings and draws conclusions. 
Some details of the methods used, 
such as the decoration transformation 
and the application of Markov chain theory, are given in the Appendices. 

\section{Potts model on a diamond chain}

\begin{figure}[h]
\includegraphics{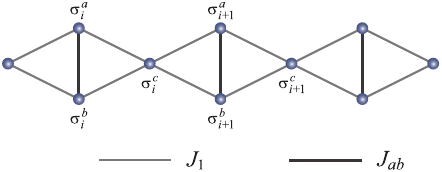}
\caption{\label{fig:Diamond-chain}
Schematic representation of the Potts model on the diamond chain structure}
\end{figure}

Despite the simplicity of the one-dimensional Potts model, it possesses
several intriguing properties that render it a worthy subject of study.
With this in mind, consider a $q$-state Potts model on a diamond
chain structure, as depicted in Fig.\ref{fig:Diamond-chain}. The
unit cell in this model is composed by three types of spins: two dimer
spins, $\sigma_{a}$ and $\sigma_{b}$, interconnected by the coupling
parameter $J_{ab}$, and a nodal spin $\sigma_{c}$, interacting with
the dimer spins through the parameter $J_{1}$. The corresponding
Potts Hamiltonian, based on this setup, can be articulated as follows:
\begin{alignat}{1}
H= & -\sum_{i=1}^{N}\left\{ J_{ab}\delta_{\sigma_{i}^{a},\sigma_{i}^{b}}
+h_{1}\delta_{\sigma_{i}^{c},1}
+h_{2}\bigl(\delta_{\sigma_{i}^{a},1}+\delta_{\sigma_{i}^{b},1}\bigr)
\right.\nonumber \\
 & +J_{1}\bigl(\delta_{\sigma_{i}^{c},\sigma_{i}^{a}}+\delta_{\sigma_{i}^{c},\sigma_{i}^{b}}+\delta_{\sigma_{i}^{c},\sigma_{i+1}^{a}}+\delta_{\sigma_{i}^{c},\sigma_{i+1}^{b}}\bigr)\Bigr\}
\label{eq:Ham1}
\end{alignat}
where $\sigma=\{1,\dots,q\}$.

It is noteworthy that the Hamiltonian \eqref{eq:Ham1} can be mapped
onto an effective one-dimensional Potts-Zimm-Bragg model~\cite{Panov},
as detailed in Appendix~\ref{sec:Dec-trans}. This suggests that the
Potts model on a diamond chain can be equated to a bona fide one-dimensional
Potts-Zimm-Bragg model as studied in reference~\cite{Panov}. It should
be noted, however, that the effective parameters of the effective
Potts-Zimm-Bragg model now depend on temperature.

\subsection{Transfer Matrix of $q$-state Potts model}

In what follows we will dedicate our attention to the thermodynamics
properties, to obtain the partition function will use the standard
transfer matrix technique. After obtaining each elements of the transfer
matrix, the $q$-dimension transfer matrix elements has the following
structure

\begin{equation}
V=\left(\begin{array}{cccccc}
d_{1} & t_{1} & t_{1} & \cdots & t_{1} & t_{1}\\
t_{1} & d_{2} & t_{2} & \cdots & t_{2} & t_{2}\\
t_{1} & t_{2} & d_{2} & \cdots & t_{2} & t_{2}\\
\vdots & \vdots & \vdots & \ddots & \vdots & \vdots\\
t_{1} & t_{2} & t_{2} & \cdots & d_{2} & t_{2}\\
t_{1} & t_{2} & t_{2} & \cdots & t_{2} & d_{2}
\end{array}\right).\label{eq:TM}
\end{equation}
Therefore, let us write the transfer matrix eigenvalues similarly
to that defined in reference~\cite{pseudo}, whose eigenvalues become

\begin{eqnarray}
\lambda_{1} & = & \frac{1}{2}\left(w_{1}+w_{-1}+\sqrt{(w_{1}-w_{-1})^{2}+4w_{0}^{2}}\right),\label{eq:Lmbd1}\\
\lambda_{2} & = & \frac{1}{2}\left(w_{1}+w_{-1}-\sqrt{(w_{1}-w_{-1})^{2}+4w_{0}^{2}}\right),\label{eq:Lmbd2}\\
\lambda_{j} & = & \left(d_{2}-t_{2}\right),\quad\text{and \ensuremath{\quad}}j=\{3,4,\dots,q\},\label{eq:Lmbdj}
\end{eqnarray}
where the elements are expressed as follow 
\begin{eqnarray}
w_{1} & = & d_{1}, \label{eq:wp1}\\
w_{-1} & = & d_{2}+(q-2)t_{2}, \label{eq:wm1}\\
w_{0} & = & \sqrt{q-1}\;t_{1},  \label{eq:w0}
\end{eqnarray}
considering the following notation
\begin{alignat}{1}
d_{1}= & z_{1}\left[\left(q-1+x^{2}z_{2}\right)^{2}+(y-1)\left(q-1+x^{4}z_{2}^{2}\right)\right],\label{eq:w1}\\
d_{2}= & \left(q-2+x^{2}+z_{2}\right)^{2}\nonumber \\
 & +(y-1)\left(q-2+x^{4}+z_{2}^{2}\right),\\
t_{1}= & \sqrt{z_{1}}\left[q-2+x\left(z_{2}+1\right)\right]^{2}\nonumber \\
 & +\sqrt{z_{1}}(y-1)\left[q-2+x^{2}\left(z_{2}^{2}+1\right)\right],\\
t_{2}= & \left(q-3+2x+z_{2}\right)^{2}\nonumber \\
 & +(y-1)\left(q-3+2x^{2}+z_{2}^{2}\right).\label{eq:t2}
\end{alignat}
here we used the following notations $x={\rm e}^{\beta J_{1}}$, $y={\rm e}^{\beta J_{ab}}$,
$z_{1}={\rm e}^{\beta h_{1}}$ and $z_{2}={\rm e}^{\beta h_{2}}$.

We can also obtain the corresponding transfer matrix eigenvectors,
which are given by 
\begin{alignat}{1}
|u_{1}\rangle= & \cos(\phi)|1\rangle+\tfrac{\sin(\phi)}{\sqrt{q-1}}\sum_{\mu=2}^{q}|\mu\rangle,\label{eq:u1}\\
|u_{2}\rangle= & -\sin(\phi)|1\rangle+\tfrac{\cos(\phi)}{\sqrt{q-1}}\sum_{\mu=2}^{q}|\mu\rangle,\label{eq:u2}\\
|u_{j}\rangle= & \sqrt{\tfrac{j-2}{j-1}}\Bigl(\tfrac{1}{j-2}\sum_{\mu=2}^{j-1}|\mu\rangle-|j\rangle\Bigr),\quad j=\{3,\cdots,q\},\label{eq:u3}
\end{alignat}
where $\phi=\frac{1}{2}\cot^{-1}\left(\frac{w_{1}-w_{-1}}{2w_{0}}\right)$,
with $-\frac{\pi}{4}\leqslant\phi\leqslant\frac{\pi}{4}$.

By using the transfer matrix eigenvalues, we express the partition
function as follows

\begin{alignat}{1}
Z_{N}= & \lambda_{1}^{N}+\lambda_{2}^{N}+(q-2)\lambda_{3}^{N},\nonumber \\
= & \lambda_{1}^{N}\left\{ 1+\Bigl(\tfrac{\lambda_{2}}{\lambda_{1}}\Bigr)^{N}+(q-2)\Bigl(\tfrac{\lambda_{3}}{\lambda_{1}}\Bigr)^{N}\right\} .\label{eq:Zn}
\end{alignat}
It is evident that the eigenvalues satisfy the following relation
$\lambda_{1}>\lambda_{2}\geqslant\lambda_{3}$. Hence, assuming $q$
finite, the free energy in thermodynamic limit ($N\rightarrow\infty$)
reduces to

\begin{equation}
f=-T\ln\left(\lambda_{1}\right).
\label{eq:FE}
\end{equation}

It is important to acknowledge that the free energy for any finite
$q$-state presents a continuous function, without any singularities
or discontinuities. As a result, we should not anticipate any genuine
phase transition at a finite temperature.

Furthermore, we can also compute the free energy~\eqref{eq:FE} from
the effective one-dimensional Potts-Zimm-Bragg model~\cite{Panov}.
The specifics of this mapping are outlined in Appendix~\ref{sec:Dec-trans}.
Note that the effective parameters of the Potts-Zimm-Bragg model are
temperature-dependent.

\section{Zero-temperature phase diagram}

In order to describe the ground state of the $q$-state Potts model
on a diamond chain we use the following notation for the state of
$i$th unit cell:
\begin{equation}
\left|\left[_{\nu_{i}}^{\mu_{i}}\right]\alpha_{i}\right\rangle_{i}
= \left\{ \left|_{\nu_{i}}^{\mu_{i}}\alpha_{i}\right\rangle_{i} 
\quad\text{or}\quad
\left|_{\mu_{i}}^{\nu_{i}}\alpha_{i}\right\rangle_{i} \right\} . 
\end{equation}
Here, $\mu_i$, $\nu_i$ and $\alpha_i$ stand for the states of sites $a$, $b$ and $c$ in the $i$th unit cell, 
and the square brackets inside a ket-vector denote two equivalent configurations for the values of Potts spins on $a$ and $b$ sites. 
Assuming $q\geqslant3$  in Hamiltonian~\eqref{eq:Ham1},
we identify the following ground states

\begin{alignat}{2}
 & \left|FM_{1}\right\rangle =\prod_{i}\left|_{1}^{1}1\right\rangle _{i}, & \quad & \left|FM_{2}\right\rangle =\prod_{i}\left|_{\mu}^{\mu}\mu\right\rangle _{i},\label{eq:wfFM1}\\
 & \left|FR_{1}\right\rangle =\prod_{i}\left|_{1}^{1}\mu_{i}\right\rangle _{i}, &  & \left|FR_{2}\right\rangle =\prod_{i}\left|\left[_{\mu_{i}}^{1}\right]1\right\rangle _{i},\label{eq:wfFM2}\\
 & \left|FR_{3}\right\rangle =\prod_{i}\left|\left[_{\nu_{i}}^{\mu_{i}}\right]\mu_{i}\right\rangle _{i}, &  & \left|FR_{4}\right\rangle =\prod_{i}\left|_{\nu_{i}}^{\nu_{i}}\mu_{i}\right\rangle _{i},\\
 & \left|FR_{5}\right\rangle =\prod_{i}\left|\left[_{\nu_{i}}^{\mu_{i}}\right]1\right\rangle _{i}, &  & \left|FR_{6}\right\rangle =\prod_{i}\left|\left[_{\nu_{i}}^{1}\right]\mu_{i}\right\rangle _{i},\\
 & \left|FR_{7}\right\rangle =\prod_{i}\left|\left[_{\nu_{i}}^{\xi_{i}}\right]\mu_{i}\right\rangle _{i} .\label{eq:wfFRs}
\end{alignat}
Here, the state indexes $\mu$, $\nu$, and $\xi$ are in a range
$2,\ldots q$, and are not equal to each other if they are written
in the same ket-vector. The cell index $i$ indicates that the site
states in neighboring cells can differ, so for the frustrated phases
the ground state consists of all relevant combinations and has non-zero
residual entropy.

\begin{figure}
\includegraphics[width=0.32\textwidth]{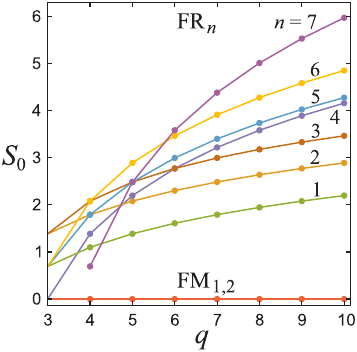} 
\caption{\label{fig:S0(q)} 
Dependencies on $q$ of the residual entropy for different phases
of the ground state (see Table~\ref{tab:GS-En-Entr-M}). }
\end{figure}

Expressions for energy and entropy per the unit cell for the ground
states~(\ref{eq:wfFM1}$-$\ref{eq:wfFRs}) are given in Table~\ref{tab:GS-En-Entr-M}.
It is important to note that the internal energy at zero temperature
does not depend on $q$, while the residual entropy is completely
determined by $q$. The frustrated phases are numbered in order of
increasing of the residual entropy for $q\geqslant7$. The dependence on
$q$ of the residual entropy for different phases is shown in Fig.\ref{fig:S0(q)}.

\begin{figure}[h]
\includegraphics[width=1\linewidth]{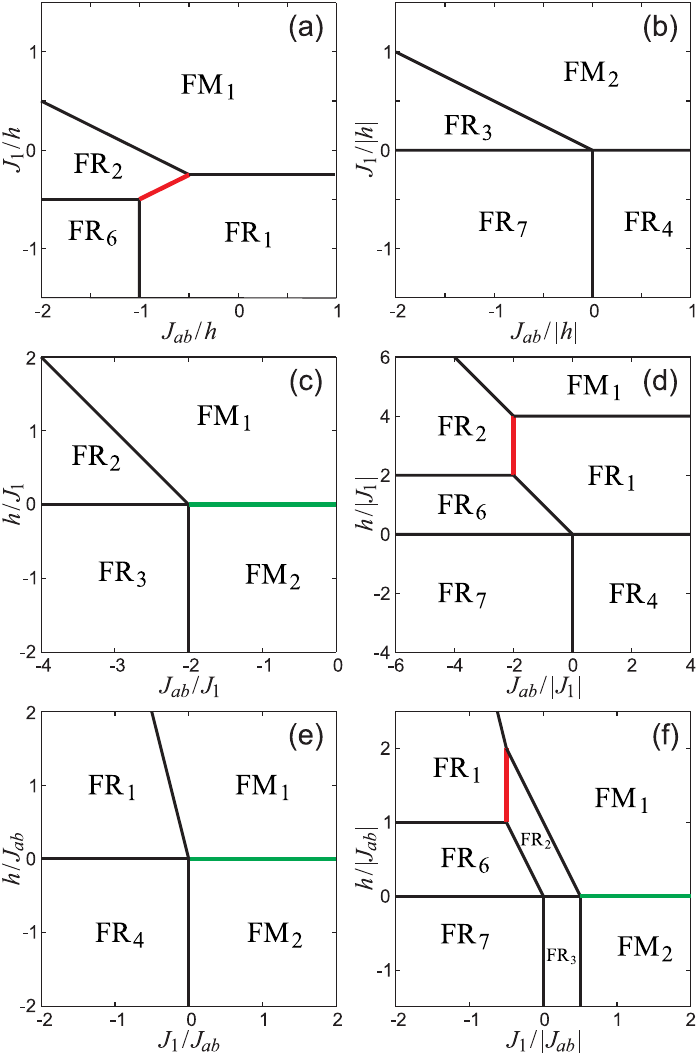} 
\caption{\label{fig:GS-q_not_3} 
The ground state phase diagrams for the case $q>3$ in the plane $J_{ab}-J_{1}$ for (a) $h>0$ and (b) $h<0$, 
in the plane $J_{ab}-h$ for (c) $J_{1}>0$ and (d) $J_{1}<0$, 
in the plane $J_{1}-h$ for (e) $J_{ab}>0$ and (f) $J_{ab}<0$. 
The green lines show the FM$_{1}$-FM$_{2}$ boundaries where both adjacent phases and the boundary phase have zero entropy, the red lines are the FR$_{1}$-FR$_{2}$ boundaries, 
where $\mathcal{S}_{{\rm FR_{1}}}<\mathcal{S}_{{\rm FR_{1}-FR_{2}}}=\mathcal{S}_{{\rm FR_{2}}}$.}
\end{figure}

The ground state phase diagrams assuming that $h_{1}=h_{2}=h$ are
shown in Fig.\ref{fig:GS-q_not_3}(a)$-$(f) in different planes.

The states FM$_{1,2}$ are of the pure ferromagnetic type. The FM$_{1}$
(FM$_{2}$) phase is realized if $h>0$ ($h<0$). In general, the
phase FM$_{2}$ is a multi-domain, and it consists of $q-1$ kinds
of equivalent macroscopic domains having all spins of a diamond chain
in the $\mu$ state. The state FR$_{1}$ is the first of frustrated
type states. The $a$ and $b$ spins are in a state $1$, while $c$-spins
may be in any of $\mu_{i}=2,\ldots q$ states, so the FR$_{1}$ phase
is realized only if $h>0$ and $J_{1}<0$. The number of states of
$c$-spins determines the entropy of the phase, $\mathcal{S}_{0}=\ln\left(q-1\right)$
per unit cell. In the second frustrated phase FR$_{2}$, the spin
$a$ equals $\mu=2,\ldots q$, and the two remaining spins in the
unit cell equal $1$, so this phase exists at $J_{ab}<0$. Due to
the equivalence of sites $a$ and $b$, the entropy of FR$_{2}$ phase
is greater by $\ln2$. Frustrated phases FR$_{3,4,7}$ exist only
if $h<0$.  
In the FR$_{3}$ phase, the spin states in the unit cell are not equal 1. 
The state of the $c$-spin and 
the state of one of the spins $a$ or $b$ are the same, $\sigma_{i}^{c}=\sigma_{i}^{(a,b)}$,
but the states of spins $a$ and $b$ in the same unit cell are different,
$\sigma_{i}^{a}\neq\sigma_{i}^{b}$, so this phase appears as a ground state only if
$J_{ab}<0$. 
Formally, the number of states of an elementary cell is $2(q-1)(q-2)$. 
But for the given phase, the state of the chain should look the same 
when moving along the chain from left to right or in the opposite direction. 
This mirror symmetry generates 
the restriction $\sigma_{i-1}^{(a,b)}=\sigma_{i}^{c}=\sigma_{i}^{(a,b)}$,
so the total number of states per unit cell in the FR$_{3}$ phase is $4(q-2)$. 
In turn, for the FR$_{4}$ phase, the conditions $\sigma_{i-1}^{(a,b)}\neq\sigma_{i}^{c}$
and $\sigma_{i}^{c}\neq\sigma_{i}^{(a,b)}$ 
must be met, so the
total number of states per unit cell reduces from $(q-1)(q-2)$ to $(q-2)^{2}$. 
Under the assumption
$h_{1}=h_{2}$, the energies of the FR$_{5}$ and FR$_{6}$ phases
are equal, and these states do not mix at the microscopic level, that
is, the unit cells of these states cannot alternate in the chain.
Formally, the chain state in the FR$_{6}$ phase region in Fig.\ref{fig:GS-q_not_3}
should be a phase separation consisting of macroscopic domains of
the FR$_{5}$ and FR$_{6}$ phases. 
The entropy of the FR$_{5}$ phase is determined by the 
total number of states in the unit cell, that is $(q-1)(q-2)$. 
In the FR$_{6}$ phase, the conditions 
$\sigma_{i-1}^{(a,b)}\neq\sigma_{i}^{c}
\neq\sigma_{i}^{(a,b)}$ give $2(q-2)^2$ states 
instead of the formally possible $2(q-1)(q-2)$ states in the unit cell. 
Nevertheless, at $q>3$ the entropy
of the FR$_{5}$ phase is less than the entropy of the FR$_{6}$ phase,
therefore, the free energy of the FR$_{6}$ phase at any finite temperature
is the lowest, and in the limit at $T\to0$ we will have the FR$_{6}$
phase as the ground state. However, the FR$_{5}$ contributes to the
state at the FR$_{2}$-FR$_{6}$ phase boundary.  
In the frustrated phase FR$_{7}$, all the spins in the unit cell are pairwise unequal
and are not equal to $1$, so, formally, the number of states of an elementary cell is $(q-1)(q-2)(q-3)$. 
The restrictions $\sigma_{i-1}^{c}\neq\sigma_{i}^{(a,b)}$ reduces the total number of states per unit cell
to the value $(q-2)(q-3)^{2}$. 
If $q\geqslant7$, the entropy of the FR$_{7}$ phase has the highest value among the other ground state phases.

\begin{figure}[h]
\includegraphics[width=1\linewidth]{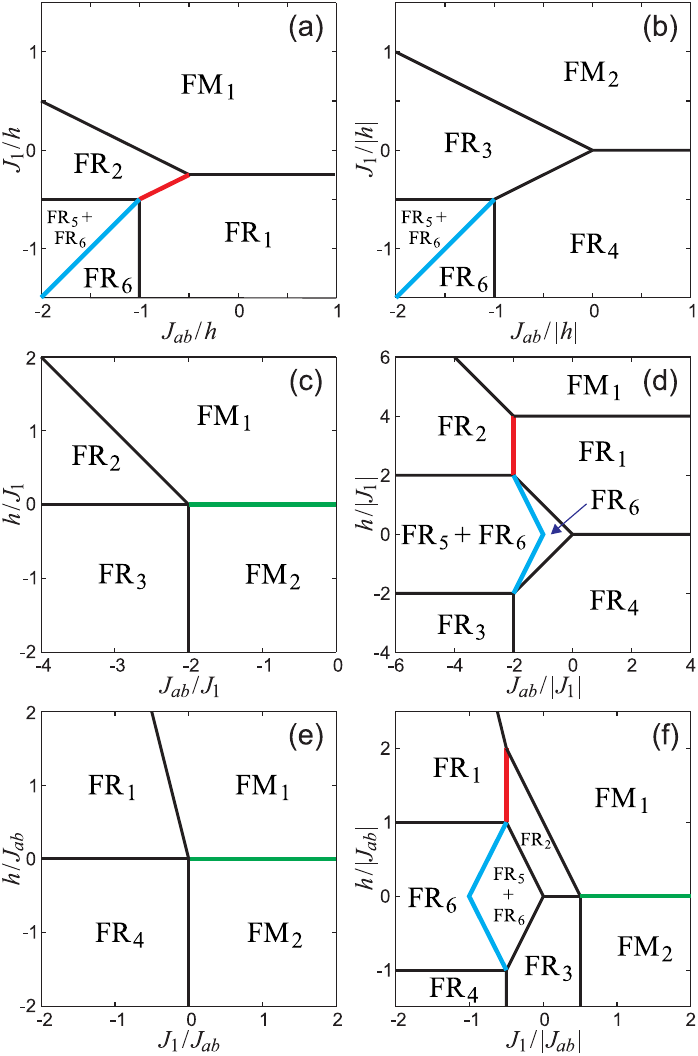} 
\caption{\label{fig:GS-q=3} 
The ground state phase diagrams for the case $q=3$ in the plane $J_{ab}-J_{1}$ for (a) $h>0$ and (b) $h<0$, 
in the plane $J_{ab}-h$ for (c) $J_{1}>0$ and (d) $J_{1}<0$, 
in the plane $J_{1}-h$ for (e) $J_{ab}>0$ and (f) $J_{ab}<0$. 
The green lines show the FM$_{1}$-FM$_{2}$ boundaries where both adjacent phases and the boundary phase have zero entropy. 
The blue lines show the new boundaries between the FR$_{6}$ phase and the phase-separated state FR$_{5}+$FR$_{6}$. 
The red lines are the FR$_{1}$-FR$_{2}$ boundaries, 
where $\mathcal{S}_{{\rm FR_{1}}}<\mathcal{S}_{{\rm FR_{1}-FR_{2}}}=\mathcal{S}_{{\rm FR_{2}}}$.}
\end{figure}

The case $q=3$ is special, and corresponding phase diagram is shown in Fig.\ref{fig:GS-q=3}. 
If $q=3$, then the states of the FR$_{7}$ phase cannot be realized, since in this case there are only 2 different Potts spins with $q\neq1$ for the 3 sites in the unit cell. 
The phase FR$_{7}$ is absent and its region on the phase diagram is taken by other phases. 
Also the phase diagram contains both the FR$_{6}$ phase and the phase-separated state ${\rm FR}_{5}+{\rm FR}_{6}$, which consist of equal fractions of the macroscopic domains of the phases FR$_{5}$ and FR$_{6}$. 
Both FR$_{6}$ and ${\rm FR}_{5}+{\rm FR}_{6}$ phases and the boundary phase have the same entropy $\mathcal{S}=\ln2$.
The structure of the ground state here can be explored using the methods of the theory of Markov chains (see Appendix~\ref{sec:Markov}).

To obtain the residual entropy $\mathcal{S}_{0}$ as a limit at zero
temperature, we use equations for the internal energy $\varepsilon$
and the free energy $f$: 
\begin{equation}
\varepsilon=f+T\mathcal{S},\qquad f=-T\ln\left(\lambda_{1}\right),
\end{equation}
where $\lambda_{1}$ is the maximum eigenvalue of the transfer matrix.
Then 
\begin{equation}
\mathcal{S}=\ln\left(\frac{\lambda_{1}}{{\rm e}^{-\beta\varepsilon}}\right).
\end{equation}
An explicit expression of $\lambda_{1}$ is given by Eq.~\eqref{eq:Lmbd1},
and we can write it in the following form 
\begin{equation}
\lambda_{1}={\rm e}^{-\beta\varepsilon_{0}}\varphi\left({\rm e}^{-\beta\left(\varepsilon_{1}-\varepsilon_{0}\right)},{\rm e}^{-\beta\left(\varepsilon_{2}-\varepsilon_{0}\right)},\ldots\right).
\end{equation}
Here $\varepsilon_{0}$ is the ground state energy for given parameters
of the Hamiltonian, so relations $\varepsilon_{k}>\varepsilon_{0}$
are fulfilled for all $k$. The form of $\varphi$ depends on the
ground state. Since $\varepsilon$ tends to $\varepsilon_{0}$ at
zero temperature, we obtain 
\begin{equation}
\mathcal{S}_{0}=\ln\left(\varphi(0)\right).\label{eq:S00}
\end{equation}
To find $\varphi(0)$, it is enough to zero out all exponential terms
having $\varepsilon_{k}\neq\varepsilon_{0}$ and replace ${\rm e}^{-\beta\varepsilon_{0}}$
by unity in $\lambda_{1}$.

A similar procedure can be defined for the magnetizations $m_{c}$
and $m_{ab}$ in the ground state. So, for the magnetizations $m_{c}$
and $m_{ab}$ we have the equations 
\begin{equation}
m_{c}=\frac{1}{\lambda_{1}}\frac{\partial\lambda_{1}}{\partial(\beta h_{1})},\quad m_{ab}=\frac{1}{\lambda_{1}}\frac{\partial\lambda_{1}}{\partial(\beta h_{2})}.
\end{equation}
If we define 
\begin{alignat}{1}
\frac{\partial\lambda_{1}}{\partial(\beta h_{1})} & ={\rm e}^{-\beta\varepsilon_{0}}\psi_{c}\left({\rm e}^{-\beta\left(\varepsilon_{1}-\varepsilon_{0}\right)},{\rm e}^{-\beta\left(\varepsilon_{2}-\varepsilon_{0}\right)},\ldots\right),\\
\frac{\partial\lambda_{1}}{\partial(\beta h_{2})} & ={\rm e}^{-\beta\varepsilon_{0}}\psi_{ab}\left({\rm e}^{-\beta\left(\varepsilon_{1}-\varepsilon_{0}\right)},{\rm e}^{-\beta\left(\varepsilon_{2}-\varepsilon_{0}\right)},\ldots\right),
\end{alignat}
then in the ground state we get 
\begin{equation}
m_{c}=\frac{\psi_{c}(0)}{\varphi(0)},\quad m_{ab}=\frac{\psi_{ab}(0)}{\varphi(0)}.\label{eq:mc00}
\end{equation}

The ground state energy $\varepsilon_{0}$, the residual entropy $\mathcal{S}_{0}$, and
magnetizations $m_{c}$ and $m_{ab}$, which were found using Equations~\eqref{eq:S00}
and~\eqref{eq:mc00}, are given in Table~\ref{tab:GS-En-Entr-M} 
for all phases and phase boundaries.

There is another way to get the values given in Table~\ref{tab:GS-En-Entr-M} 
and study the properties of the ground state in detail.
This method, based on the theory of Markov chains, is described in Appendix~\ref{sec:Markov}.

\begin{table*}[thb]
\caption{\label{tab:GS-En-Entr-M} 
The ground state energy, residual entropy
and magnetizations of a diamond Potts chain.}
{ 
\global\long\def\arraystretch{1.4}%
\begin{ruledtabular}
\begin{tabular}{cllll}
Ground state & $\varepsilon_{0}$  & $\mathcal{S}_{0}$  & $m_{c}$  & $m_{ab}$\tabularnewline
\hline 
FM$_{1}$  & $-\left(4J_{1}+J_{ab}+3h\right)$  & $0$  & $1$  & $2$ \tabularnewline
FM$_{2}$  & $-\left(4J_{1}+J_{ab}\right)$  & $0$  & $0$  & $0$ \tabularnewline
FR$_{1}$  & $-\left(J_{ab}+2h\right)$  & $\ln\left(q-1\right)$  & $0$  & $2$ \tabularnewline
FR$_{2}$  & $-\left(2J_{1}+2h\right)$  & $\ln\left[2(q-1)\right]$  & $1$  & $1$ \tabularnewline
FR$_{3}$  & $-2J_{1}$  & $\ln\left[4(q-2)\right]$  & $0$  & $0$ \tabularnewline
FR$_{4}$  & $-J_{ab}$  & $2\ln(q-2)$  & $0$  & $0$ \tabularnewline
FR$_{5}$  & $-h$  & $\ln\left[(q-1)(q-2)\right]$  & $1$  & $0$ \tabularnewline
FR$_{6}$  & $-h$  & $\ln\left[2(q-2)^{2}\right]$  & $0$  & $1$ \tabularnewline
FR$_{7}$  & $0$  & $\ln\left[(q-2)(q-3)^{2}\right]$  & $0$  & $0$ \tabularnewline
\hline 
FM$_{1}$-FM$_{2}$  & $-\left(4J_{1}+J_{ab}\right)$  & $0$  & $\frac{1}{q}$  & $\frac{2}{q}$\tabularnewline
FM$_{1}$-FR$_{1}$  & $8J_{1}-J_{ab}$  & $\ln(q)$  & $\frac{1}{q}$  & 2\tabularnewline
FR$_{1}$-FR$_{2}$  & $-2\left(J_{1}+h\right)$  & $\ln[2(q-1)]$  & 1  & 1\tabularnewline
FM$_{1}$-FR$_{2}$  & $2\left(J_{1}+J_{ab}\right)$  & $\ln(2q-1)$  & 1  & $\frac{2q}{2q-1}$\tabularnewline
FM$_{2}$-FR$_{3}$  & $-2J_{1}$  & $\ln(4q-7)$  & 0  & 0\tabularnewline
FR$_{2}$-FR$_{3}$  & $-2J_{1}$  & $\ln[4(q-1)]$  & $\frac{1}{q}$  & $\frac{2}{q}$\tabularnewline
FM$_{2}$-FR$_{4}$  & $-2J_{ab}$  & $2\ln(q-1)$  & 0  & 0\tabularnewline
FR$_{1}$-FR$_{4}$  & $-J_{ab}$  & $2\ln(q-1)$  & $\frac{1}{q}$  & $\frac{2}{q}$\tabularnewline
FR$_{1}$-FR$_{6}$  & $J_{ab}$  & $\ln(2q^{2}-7q+7)$  & 0  & $\frac{2(q^{2}-3q+3)}{2q^{2}-7q+7}$\tabularnewline
FR$_{2}$-FR$_{6}$  & $2J_{1}$  & $\ln\left[\frac{1}{2}\left(3q^{2}-9q+8+\phi_{1}(q)\right)\right]$\footnote{$\phi_{1}(q)=\sqrt{q\left(q^{3}+2q^{2}-15q+16\right)}$}  & $\frac{\phi_{1}(q)-q^{2}+7q-8}{2\phi_{1}(q)}$  & $\frac{\phi_{1}(q)+q^{2}-3q+4}{2\phi_{1}(q)}$\tabularnewline
FR$_{4}$-FR$_{7}$  & $0$  & $\ln[(q-2)(q^{2}-5q+7)]$  & 0  & 0\tabularnewline
FR$_{6}$-FR$_{7}$  & $0$  & $\ln[(q-1)(q-2)^{2}]$  & $\frac{1}{q}$  & $\frac{2}{q}$\tabularnewline
FR$_{3}$-FR$_{7}$  & $0$  & $\ln[(q-1)^{2}(q-2)]$  & 0  & 0\tabularnewline
\hline 
FM$_{1}$-FR$_{1}$-FR$_{2}$  & $6J_{1}$  & $\ln\left[\frac{1}{2}\left(3q-2+\phi_{2}(q)\right)\right]$\footnote{$\phi_{2}(q)=\sqrt{q^{2}+4q-4}$}  & $\frac{q+\phi_{2}(q)}{2\phi_{2}(q)}$  & $\frac{3\phi_{2}(q)-q+2}{2\phi_{2}(q)}$\tabularnewline
FR$_{1}$-FR$_{2}$-FR$_{6}$  & $2J_{1}$  & $\ln\left[\frac{1}{2}\left(3q^{2}-8q+7+\phi_{3}(q)\right)\right]$\footnote{$\phi_{3}(q)=\sqrt{q^{4}+4q^{3}-30q^{2}+44q-15}$}  & $\frac{\phi_{3}(q)-q^{2}+6q-7}{2\phi_{3}(q)}$  & $\frac{2\left[\left(q^{2}-2q+2\right)\left(q^{2}-7+\phi_{3}(q)\right)+10-2q\right]}{\phi_{3}(q)\left(3q^{2}-8q+7+\phi_{3}(q)\right)}$\tabularnewline
FM$_{1}$-FM$_{2}$-FR$_{2}$-FR$_{3}$ & $-2J_{1}$  & $\ln(4q-3)$  & $\frac{1}{q}$  & $\frac{2}{q}$\tabularnewline
FR$_{1}$-FR$_{4}$-FR$_{6}$-FR$_{7}$ & $0$  & $\ln\left[\left(q-1\right)\left(q^{2}-3q+3\right)\right]$  & $\frac{1}{q}$  & $\frac{2}{q}$\tabularnewline
FM$_{2}$-FR$_{3}$-FR$_{4}$-FR$_{7}$ & $0$  & $3\ln(q-1)$  & 0  & 0\tabularnewline
FR$_{2}$-FR$_{3}$-FR$_{6}$-FR$_{7}$ & $0$  & $\ln[q^{2}(q-1)]$  & $\frac{1}{q}$  & $\frac{2}{q}$\tabularnewline
O  & $0$  & $3\ln(q)$  & $\frac{1}{q}$  & $\frac{2}{q}$\tabularnewline
\end{tabular}
\end{ruledtabular}
} 
\end{table*}

The values in Table~\ref{tab:GS-En-Entr-M} show that the entropy of all
phase boundaries is greater than the entropy of adjacent phases. 
The exceptions are two phase boundaries. 
The first is the FM$_{1}$-FM$_{2}$ boundary, 
where the entropy of both adjacent phases and the boundary state is zero. 
The second is the FR$_{1}$-FR$_{2}$ boundary, 
where the boundary state is such that 
$\mathcal{S}_{{\rm FR_{1}}}<\mathcal{S}_{{\rm FR_{1}-FR_{2}}}=\mathcal{S}_{{\rm FR_{2}}}$.
This phase boundary is truly an anomalous property, 
leading to a peculiar phase pseudo-transition at finite temperature, 
which we will explore in the next section.

\section{Thermodynamics of $q$-state Potts model}

In what follows, we will analyze the thermodynamic properties of the model in detail. First, we will examine the 3-state models ($q = 3$), which exhibit some peculiar properties, distinct from the behavior for $q > 3$. Later, we will explore the case when $q > 3$. It's worth noting that the behavior for any $q > 3$ tends to be rather consistent across finite values of $q$. For the purpose of this discussion, we will focus specifically on $q = 5$, without losing its core properties.

\subsection{3-state Potts Model}
Indeed, the 2-state Potts model is equivalent to the Ising model, which differs significantly from the $q>2$ state Potts model. A primary feature to highlight in the latter is the emergence of frustration. The 3-state Potts model, being the first to exhibit this frustration behavior, is expected to display peculiar characteristics. 
In contrast, all higher $q$-state Potts models tend to behave similarly. 
The one-dimensional 3-state Potts model, provides a richer set of
behaviors than the 2-state Ising model, yet it is still analytically
solvable. 
Surely, in the one-dimensional case, there is no phase transition
at finite temperature for the Potts model with $q>2$ states, which
can be proven via the Peierls argument~\cite{baxter2016exactly}.
This property makes the 1D 3-state Potts model a tractable system
to study, helping investigations into more intricate systems and behaviors
within statistical physics. Its study contributes to the broader field
of statistical physics and has implications in several scientific
disciplines. In this sense here we will consider the special case
of $3$-state Potts model on diamond chain.

Initially, we will explore the thermodynamics and magnetization properties 
in the vicinity of the zero-temperature phase boundary that separates 
FM$_{1}$ and FM$_{2}$. 

\begin{figure}
\includegraphics[width=0.45\textwidth]{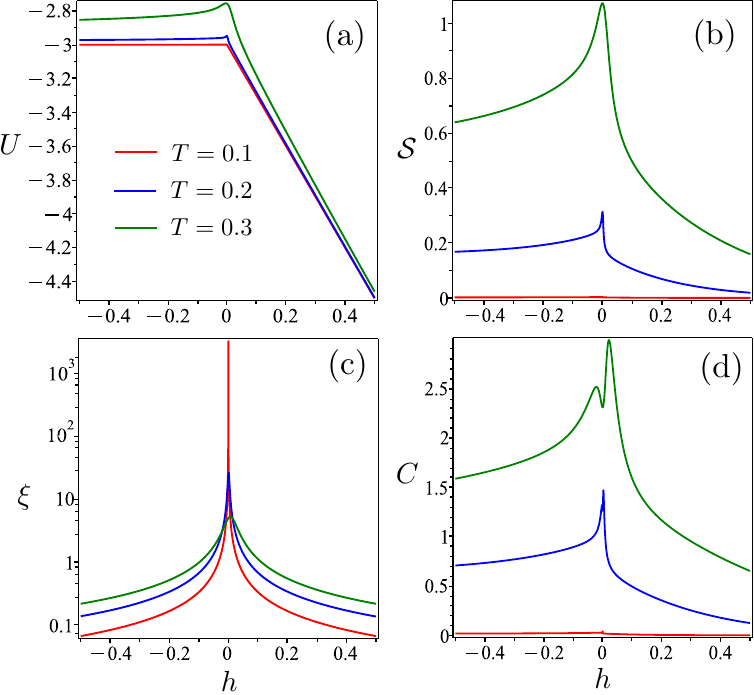}
\caption{\label{fig:SCUx}(a) 
Internal energy $U$ as a function of the external magnetic field, 
for three different temperature values $T=\{0.1,0.2,0.3\}$ assuming 
fixed parameter $J_{ab}=-1$, $J_{1}=1$ and $q=3$; (b) the entropy $\mathcal{S}$ 
for the same conditions; (c) the correlation length, and (d) the specific 
heat.}
\end{figure}

Figure \ref{fig:SCUx}a illustrates the internal energy $U$ as a
function of the external magnetic field. We assume the same magnetic
field for both nodal and dimer sites ($h_{1}=h_{2}=h$). Three different
temperature values are considered to demonstrate the behavior of the
internal energy. At $h=0$, corresponding to the zero-temperature
phase transition between FM$_{1}$ and FM$_{2}$ (refer to Fig.\ref{fig:GS-q=3}),
an evident change is observed. As the temperature increases, a small
peak emerges at $h=0$, which grows with higher temperatures. In panel
(b), the entropy $\mathcal{S}$ is shown as a function of the external
magnetic field, under the same conditions as panel (a). Here, we notice
the absence of residual entropy at zero temperature, in accordance
with the argument in reference \cite{Rojas2020-1,Rojas2020-2}, suggesting the presence
of a pseudo-critical temperature at this boundary. Panel (c) displays
the correlation length  
$\xi=1/\ln\left(\tfrac{\lambda_{1}}{\lambda_{2}}\right)$ 
as a function of $h$, using the same
parameter set as the previous panels. Once again, a sharp peak at
$h=0$ confirms the phase transition at zero temperature. Interestingly,
in this case, there are not only one pseudo-critical temperature but
infinitely many. For any temperature $T_{p}\lesssim0.2$, we observe
a sharp peak in the correlation length at a null magnetic field. Lastly,
panel (d) presents the specific heat under the same conditions. In
contrast to a typical pseudo-critical peak, an intense peak appears,
with a small minimum at $h=0$. As the temperature decreases, the
specific heat tends to zero, as expected.

\begin{figure}
\includegraphics[width=0.45\textwidth]{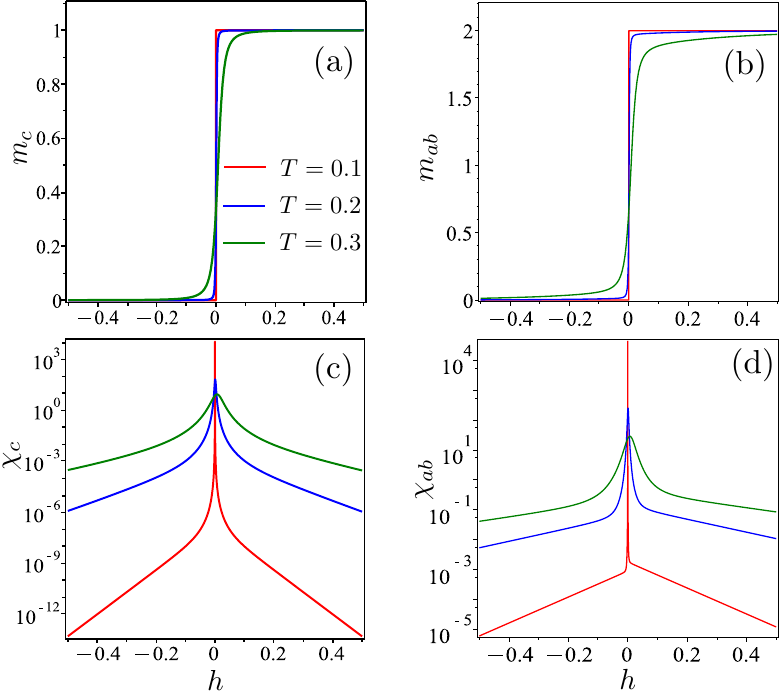}
\caption{\label{fig:MgXs}(a) 
Magnetization $m_{c}$ as a function of external
magnetic field $h$, for three different temperature values $T=\{0.1,0.2,0.3\}$,
assuming fixed parameter  $J_{ab}=-1$, $J_{1}=1$ and $q=3$. (b) Magnetization
$m_{ab}$ as a function of external magnetic field $h$, for the same
set of fixed parameters in panel (a). (c-d) Magnetic susceptibility
$\chi_{c}$ and $\chi_{ab}$ assuming same condition the above panels. }
\end{figure}

In the following analysis, we explore the magnetic properties of the
system, specifically the magnetization and magnetic susceptibility.
Fig. \ref{fig:MgXs}a illustrates the magnetization $m_{c}$ of the
nodal site as a function of the external magnetic field $h$ for temperatures
$T=\{0.1,0.2,0.3\}$. The parameters $J_{ab}=-1$ and $J_{1}=1$ remain
fixed throughout. In the low-temperature region, we observe the saturated
phase (FM$_{1}$) and a phase transition at $h=0$, where the magnetization
drops to zero, corresponding to FM$_{2}$. It is important to note
that FM$_{2}$ exhibits null magnetization since, according to the
definition in Eq.~\eqref{eq:wfFM2}, it aligns in any state other
than $1$. Moving on to panel (b), we present the dimer magnetization
$m_{ab}$, which exhibits a similar behavior to panel (a). Panel (c)
exhibits the magnetic susceptibility $\chi_{c}$ as a function of
the external magnetic field $h$, under the same aforementioned conditions.
Notably, it displays sharp peaks reminiscent of pseudo-critical phase
transitions, particularly for temperatures $T_{p}\lesssim0.2$. Similarly,
panel (d) illustrates the dimer magnetic susceptibility $\chi_{ab}$
as a function of $h$, employing the same conditions as the previous
panels. These last two panels exhibit characteristic sharp peaks distinct
from the double peak observed in the specific heat plot depicted in
Fig.\ref{fig:SCUx}d, which occurs around $h=0$.

There is a peculiar behavior for $q=3$,   so we will now investigate the anomalous interface between FR$_{5}$ and FR$_{5}+$FR$_{6}$, which represents another phase boundary that
requires analysis.

\begin{figure}
\includegraphics[width=0.45\textwidth]{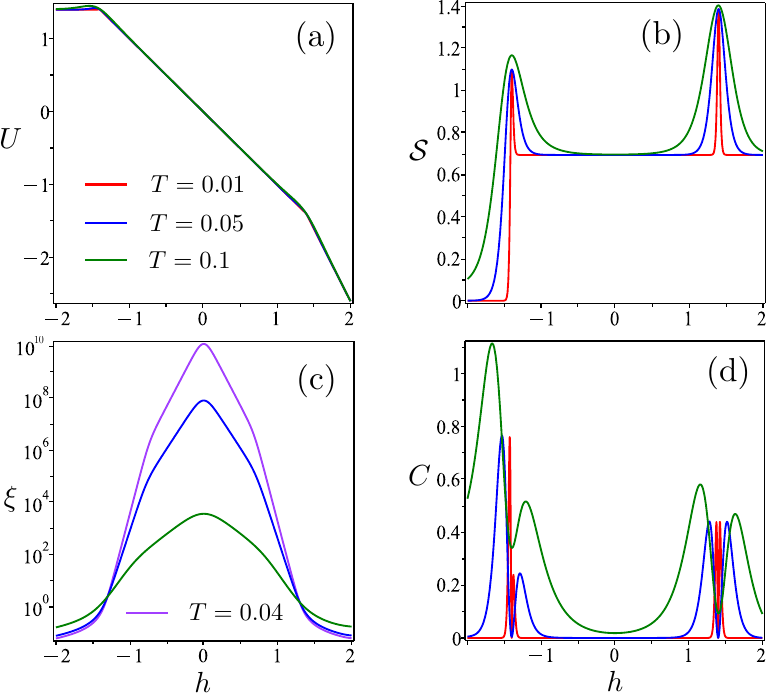}
\caption{\label{fig:SCUc}(a) 
Internal energy $U$ as a function of temperature for three specific values ($T=\{0.01,0.05,0.1\}$), assuming fixed
parameters $J_{ab}=-1.4$ ,  $J_{1}=-1$ and $q=3$; (b) the entropy $\mathcal{S}$
under the same conditions; (c) the correlation length $\xi$; and
(d) the specific heat $C$.}
\end{figure}

Figure \ref{fig:SCUc}a illustrates the internal energy ($U$) as
a function of the external magnetic field ($h$), with parameters
$J_{ab}=-1.4$ and $J_{1}=-1$ held constant. The graph uses three distinct
temperatures for illustrative purposes. The internal energy for these
temperatures appears almost identical, with minor variations around
$h=\pm1.4$, where a zero-temperature phase transition occurs. Conversely,
panel (b) depicts the entropy ($\mathcal{S}$) as a function of the
same set of temperatures. Unlike the internal energy, the entropies
for these temperatures are distinctly different under the same set
of parameters. A peak is noticeable at the same magnetic field $h=\pm1.4$,
underscoring the impact of the zero-temperature phase transition.
On the other hand, the correlation length ($\xi$) indicates a curvature
change at a disparate temperature, approximately around $h\approx\pm0.8$,
but no evidence of phase transition influence at $h=\pm1.4$ is discernible.
Lastly, panel (d) demonstrates the specific heat as a function of
temperature, maintaining the same parameters as in panel (a). A double
peak is observable around $h=\pm1.4$, but no signs of unusual behavior
are evident at $h\approx\pm0.8$. 
Although there is no anomalous behavior for $U$, $\cal S$, and $C$ at $h=0$, 
the correlation length $\xi$ illustrates a maximum at $h=0$. 
This anomalous behavior will be discussed further later.

\begin{figure}
\includegraphics[width=0.45\textwidth]{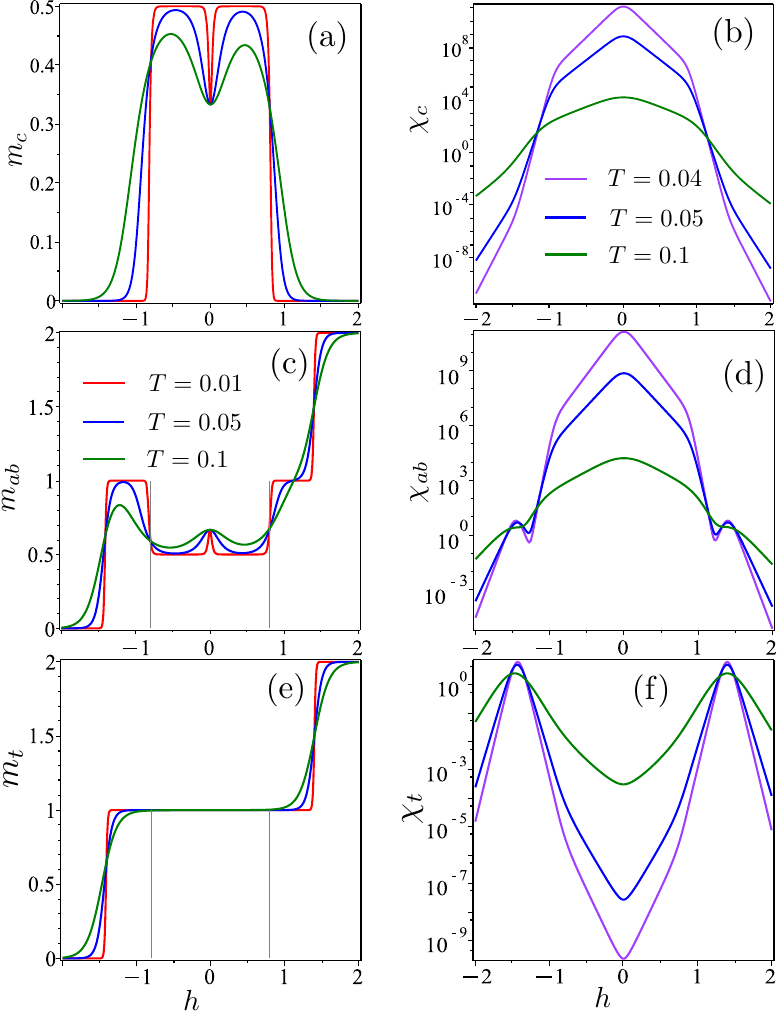}
\caption{\label{fig:MgXsFr} 
(a) Magnetization $m_{c}$ as a function of the
external magnetic field $h$, considering three different temperature
values $T=\{0.01,0.05,0.1\}$. The fixed parameters $J_{ab}=-1.4$, $J_{1}=-1$ and $q=3$  are assumed. 
(b) Magnetization $m_{ab}$ as a function of the external magnetic field $h$, 
with the same set of fixed parameters as in panel (a). 
(c) Magnetic susceptibility $\chi_{c}$ and (d) $\chi_{ab}$ 
under the same conditions as the previous panels.}
\end{figure}

Figure \ref{fig:MgXsFr}a depicts the magnetization of nodal spin, assuming
parameters set in Fig.\ref{fig:SCUc}. It reveals that the magnetization,
denoted as $m_{c}$, alters its behavior notably at $h\approx\pm0.8$.
However, there are no traces of a phase transition at $h=\pm1.4$,
even though the magnetization exhibits symmetry under the exchange
of the magnetic field sign.  Additionally, panel (b) reports the magnetic susceptibility $\chi_{c}$
of nodal spin, as a function of the magnetic field, with temperatures
designated in the same panel. Note that, the magnetic susceptibility
enlarges rapidly for lower temperatures and remains substantial at
$h=\pm1$. It also displays a significant alteration in the curve
around $h\approx\pm0.8$ and a change in curvature at about $h\approx\pm1.4$. 
In contrast, the panel (c) illustrates the dimer magnetization,
denoted as $m_{ab}$, has been represented as a function of the external
magnetic field $h$. Here, we observe the zero-temperature phase transition
impact at $h=\pm1.4$ and $h=\pm0.8$, despite the magnetization no
longer maintaining symmetry under the exchange of the magnetic field.
Similarly, panel (d) features the magnetic susceptibility $\chi_{ab}$,
using the same parameters presented in panel (a). Comparable to observations
in panel (c), we detect a significant change of curvature around $h\approx\pm0.8$,
while a local maximum of magnetic susceptibility emerges at $h\approx\pm1.4$. 
As previously identified in the correlation length $\xi$, there is an anomalous behavior observed at $h=0$. The magnetization $m_c$ exhibits a peculiar value of $1/3$ at null magnetic field, while similarly, $m_{ab}$  yields $2/3$ at $h=0$, and obviously the total magnetization becomes 1. This anomalous behavior is also manifested in the magnetic susceptibilities $\chi_c$ and $\chi_{ab}$, which exhibit a maximum value at $h=0$ when the magnetic field is varied.
Furthermore, in panel (e), we report the total magnetization \(m_t = m_c + m_{ab}\). Interestingly, based on our observations, there is no evidence of any anomalous behavior; instead, a long plateau is evident. For this analysis, we assumed the same set of parameters as those used for the previous partial magnetizations. 
Panel (f) shows the total magnetic susceptibility \(\chi_t = \chi_c + \chi_{ab} + 2\chi_{abc}\), where \(\chi_{abc} = -\frac{\partial^2 f}{\partial h_c\partial h_{ab}}\) (not depicted). Again, we relied on the parameters established for the partial magnetic susceptibilities. It is noteworthy that this analysis does not reveal significant insights around the anomalous regions. Additionally, the total magnetic susceptibility presents a markedly smaller magnitude compared to the partial magnetic susceptibilities displayed in panels (b) and (d). This reduced magnitude arises because the magnetic susceptibility \(\chi_{abc}\) counterbalances the positive contributions from \(\chi_c\) and \(\chi_{ab}\), due to its comparable magnitude. As an alternative approach, one can determine \(\chi_t\) for the current case by assuming \(h_c = h_{ab}\) and taking the second derivative of the negative free energy. 

\subsection{Pseudo-critical temperature around FR$_{1}-$FR$_{2}$ phase boundary}

We will now investigate the properties of the Potts model on a diamond
structure, which displays anomalous behavior near the phase boundary
FR$_{1}-$FR$_{2}$ influenced by temperature variations. This region
displays a pseudo-critical transition, akin to a first or second-order
phase transition. Notably, the anomalous properties observed in the
low-temperature regime are primarily independent of the particular
value of $q$.  For $q > 3$, the behavior of physical quantities is rather similar. Therefore, we will consider $q = 5$ solely for illustrative purposes, without losing any relevant properties.  Examining this transition is crucial for comprehending the physical properties of the Potts model and predicting its behavior under diverse conditions.

\begin{figure}
\includegraphics[width=0.45\textwidth]{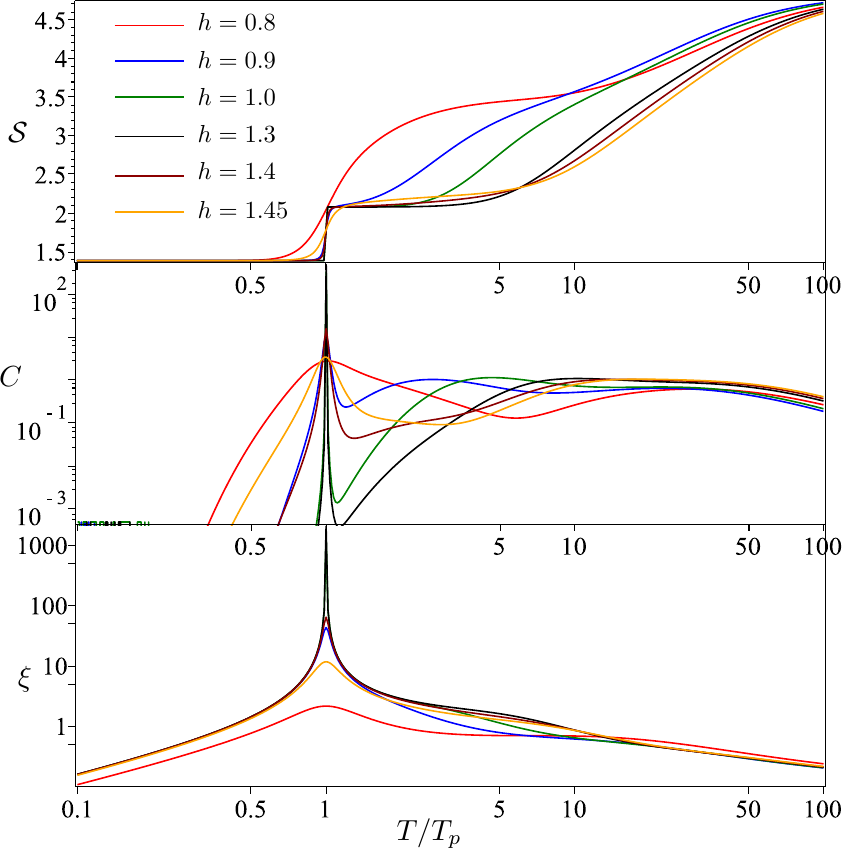}
\caption{\label{fig:SCxi}
Entropy $\mathcal{S}$ (top), specific heat $C$
(middle), and correlation length $\xi$ (bottom), as function of temperature
$T/T_{p}$, in units of pseudo-critical temperature $T_{p}$ assuming
fixed parameter $J_{ab}=-0.75$, $J_{1}=-0.38$ and $q=5$, for several
magnetic field $h=\{0.8,0.9,1.0,1.3,1.4,1.45\}$, and corresponding
pseudo-critical temperature $T_{p}=\{$$0.0100651534871$, $0.0143473088244$,
$0.0144268117883$, $0.0144268970579$, $0.0143982497686$, $0.0139830161395\}$,
respectively.}
\end{figure}

\subsubsection{Entropy }

Fig.\ref{fig:SCxi}(top) shows the plot of entropy ($\mathcal{S}$)
as a function of temperature ($T/T_{p}$), where $T_{p}$ is the pseudo-critical
temperature, with $J_{ab}=-0.75$, $J_{1}=-0.38$ and $q=5$ are fixed
parameters. We consider several magnetic fields, i.e., $h=\{0.8,0.9,1.0,1.3,1.4,1.45\}$,
and their corresponding pseudo-critical temperatures are $T_{p}=\{$$0.0100651534871$,
$0.0143473088244$, $0.0144268117883$, $0.0144268970579$, $0.0143982497686$,
$0.0139830161395\}$, respectively. For magnetic fields in the range
of $1\lesssim h\lesssim1.3$, we observe a robust change of curvature
at $T_{p}$, which resembles a typical first-order phase transition.
However, there is no sudden jump in entropy at $T_{p}$, and when
we magnify the entropy plot around $T_{p}$, we can see that the curve
is a continuous smooth function. On the other hand, for other magnetic
field values, the sudden change of curves is clearly a smooth function
(not shown). It is worth mentioning that for $T<T_{p}$, the system
mostly resembles the FR$_{1}$ phase with residual entropy $\mathcal{S}=\ln(q-1)=\ln(4)$,
while for $T>T_{p}$, the system behaves somewhat similarly to the
FR$_{2}$ phase, with residual entropy $\mathcal{S}\approx\ln(2(q-1))=\ln(8)$.
This effect is more evident for magnetic fields in the range of $1\lesssim h\lesssim1.3$.

\subsubsection{Specific heat}

In Fig.\ref{fig:SCxi}(middle), we plot specific heat ($C$) as a
function of temperature ($T/T_{p}$), or in units of pseudo-critical
temperature $T_{p}$. We consider the same set of parameters as in
the previous plot, and each colored curve corresponds to the caption
specified in the top panel. The anomalous behavior manifests clearly
for magnetic fields in the range of $1\lesssim h\lesssim1.3$, where
we observe a very intense sharp peak around $T_{p}$ or $T/T_{p}=1$,
which looks like a second-order phase transition. However, there is
no divergence at $T_{p}$. For other values of magnetic field, this
peak becomes broader and less intense. This sharp peak around $T_{p}$
evidently signals the limit between the FR$_{1}$ phase and FR$_{2}$
phase, as discussed earlier. Therefore, the plots in Fig.\ref{fig:SCxi}
provide valuable insights into the magnetic field-induced phase transition
in the system.

\subsubsection{Correlation Length}

In Fig.\ref{fig:SCxi}(bottom), we plot the correlation length ($\xi$)
as a function of temperature ($T/T_{p}$), [in units of pseudo-critical
temperature $T_{p}$]. For simplicity and consistency with the previous
figures, we consider the same set of parameters as in the top panel.
Again, we observe the anomalous behavior of the correlation length
around $T_{p}$, confirming the evidence of a pseudo-transition at
$T_{p}$. The correlation length peak is more intense when we consider
an external magnetic field in the range of $1\lesssim h\lesssim1.3$.
This peak originates when the second largest eigenvalue becomes as
important as the largest eigenvalue, although it should never attain
the magnitude of the largest eigenvalue. For other values of magnetic
field, the peak becomes less intense. These results further support
the evidence of a magnetic field-induced phase transition in the system,
as seen in the previous plots of specific heat and entropy. The behavior
of the correlation length also provides valuable insights into the
nature of this transition.

The power-law behavior of the correlation length may
be analytically derived using the formula proposed in reference \cite{Univrslt}.
This can be achieved by manipulating the relation: $\xi=1/\ln\left(\tfrac{\lambda_{1}}{\lambda_{2}}\right)$.
Utilizing the effective Boltzmann factors from \eqref{eq:wp1} and
\eqref{eq:wm1}, we can express the correlation length as
\begin{equation}
\xi(\tau)=c_{\xi}|\tau|^{-1}+{\cal O}(\tau^{2}),
\end{equation}
where 
\begin{equation}
c_{\xi}=\frac{1}{\tilde{w}_{1}T_{p}}\left|\tfrac{\partial\left[w_{1}(\beta)-w_{-1}(\beta)\right]}{\partial\beta}\right|_{\beta=\beta_{p}} , 
\end{equation}
and $\tau=(T_{p}-T)/T_{p}$ with $\tilde{w}_{1}=w_{1}(\beta_{p})$.

\begin{figure}
\includegraphics[width=0.47\textwidth]{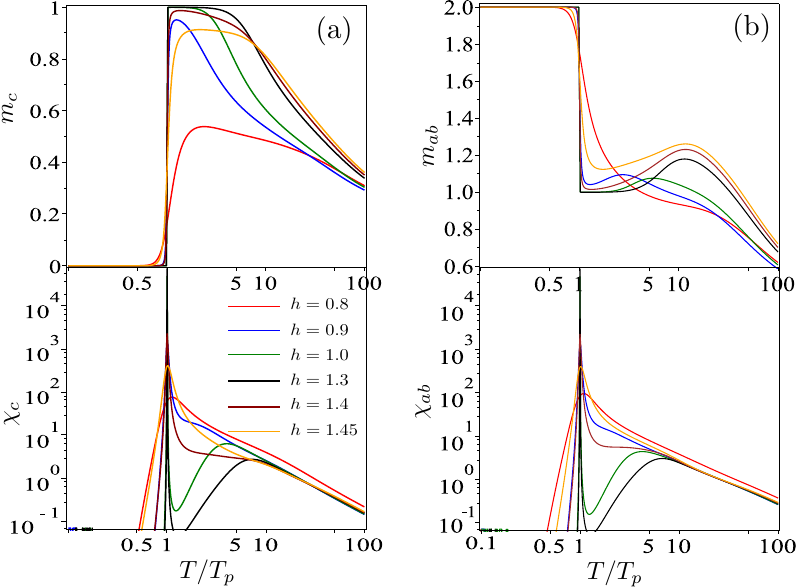}
\caption{\label{fig:MX}
Magnetization, Magnetic susceptibility as a function 
of temperature $T/T_{p}$, in units of pseudo-critical temperature 
$T_{p}$, assuming the values considered in Fig.\ref{fig:SCxi}. (a) 
Corresponds to the magnetization of nodal spins $m_{c}$ (top) and 
corresponding magnetic field (bottom).}
\end{figure}

\subsubsection{Magnetization}

In Fig.\ref{fig:MX}a (top), we plot the magnetization $m_{c}$ as
a function of temperature $T/T_{p}$, in units of pseudo-critical
temperature $T_{p}$. We consider the same fixed parameter set as
in Fig.\ref{fig:SCxi} for comparison purposes. It is evident that
for $T<T_{p}$, the magnetization $m_{c}$ is almost negligible, indicating
that almost none of the spin components are in the first component
of spin. Suddenly, the magnetization increases rapidly and reaches
a saturated value at $T/T_{p}\approx4$, indicating that the spins
are almost fully ordered. For higher temperatures, the spins gradually
become randomly oriented. This behavior is more pronounced for the
magnetic field range of $1\lesssim h\lesssim1.3$, while for other
values of magnetic field, the magnetization $m_{c}$ shows a smooth
curve with an enhanced magnetization slightly above $T_{p}$. Similarly,
in Fig.\ref{fig:MX}b (top), we present the magnetization $m_{ab}$
as a function of temperature in units of $T_{p}$. In this case, the
magnetization $m_{ab}$ is well-behaved, and most of the particles
spin components are configured in the FR$_{1}$ phase. For $1\lesssim T/T_{p}\lesssim4$,
the spin of the system is roughly configured in the FR$_{2}$ phase,
and then increases slightly. However, this peak disappears when the
magnetic field satisfies the condition of $h\lesssim1$ and $h\gtrsim1.3$.
As the temperature increases further, the magnetization decreases
asymptotically.

\subsubsection{Magnetic Susceptibility}

In Fig.\ref{fig:MX}a (bottom), we present the nodal spin magnetic
susceptibility $\chi_{c}$ as a function of temperature $T/T_{p}$,
where we use the same set of parameters as in the above panels for ease of comparison. In the range of magnetic field $1\lesssim h\lesssim1.3$,
the $\chi_{c}$ peak is very sharp around $T/T_{p}=1$, and a second
broader peak appears at higher temperatures, which vanishes when the
peak at $T_{p}$ decreases. For other intervals of magnetic field,
the magnetic susceptibility exhibits less intense and broader peaks
around $T/T_{p}\approx1$, and when the peak becomes less pronounced,
the second peak disappears as well. Similarly, in Fig.\ref{fig:MX}b
(bottom), we report the magnetic susceptibility $\chi_{ab}$ as a
function of temperature $T/T_{p}$. For magnetic fields $1\lesssim h\lesssim1.3$, the intense
sharp peak delimits the boundary between quasi-phases $qFR_{1}$ and
$qFR_{2}$, accompanied by a second broader peak at higher temperatures.
However, for other intervals of magnetic field, the intense sharp
peak decreases and gradually disappears, and at the same time, the
second broad peak vanishes as well.

To summarize our results, it is important to note that while the transfer matrix of most models exhibiting pseudo-transitions is typically reduced to a \(2 \times 2\) matrix as shown in reference \cite{Univrslt}, our transfer matrix can, in principle, be significantly larger, depending on the values of \(q\). This contrasts with what was previously discussed in \cite{Univrslt}. However, both the largest and the second-largest eigenvalues share the same structure as those of a typical \(2 \times 2\) transfer matrix. In the thermodynamic limit, all other eigenvalues become irrelevant. Therefore, it's worth mentioning that pseudo-transitions adhere to the same universality properties outlined in reference \cite{Univrslt}.

\section{Conclusions}

Here we explored the $q$-state Potts model on a diamond chain in
order to study the zero-temperature phase transitions and thermodynamic properties.
The $q$-state Potts model on a diamond chain exhibits intriguing behavior,
due to discrete states assembled on a diamond chain structure, which
exhibits several unusual features, like various possible alignments
of magnetic moments.

The 3-state Potts model on a diamond chain presents peculiar characteristics,
around the zero temperature phase transition FM$_{1}-$FM$_{2}$
and FR$_{5}$ and FR$_{5}+$FR$_{6}$, such as the absence of residual
entropy at the phase boundary. Thermodynamic quantities such as entropy,
internal energy, and specific heat remain unaffected by this phase
transition, even at significantly low temperatures, while magnetic
properties like correlation length, magnetization, and magnetic susceptibility
offer evidence of a zero-temperature phase transition at finite temperatures
when the magnetic field is varied. These findings highlight the intricate
nature of the $q$-state Potts model on a diamond chain and contribute
to our understanding of complex systems in diverse scientific disciplines.

Furthermore, we conducted an analysis of the $q$-state Potts model, primarily independent of the specific value of $q$, but for illustrative purposes, we chose $q=5$. Our exploration centered around the phase boundaries FR$_{1}$ and FR$_{2}$, where certain anomalous properties become more pronounced in low-temperature regions. This is due to residual
entropy, which unveils unusual frustrated regions at zero-temperature
phase transitions. Phase boundaries featuring non-trivial phase transitions
demonstrate anomalous thermodynamic properties, including a sharp
entropy alteration as a function of temperature, resembling a first-order
jump of entropy without an actual discontinuity. 
Similarly, second-order 
derivatives of the free energy, such as specific heat and magnetic
susceptibility, present distinct peaks akin to those found in second-order
phase transition divergences, but without any singularities. The correlation
length also exhibits analogous behavior at the pseudo-critical temperature,
marked by a sharp and robust peak that could be easily misinterpreted
as true divergence.
It is worth noting that, although the ground state phase diagram 
shows several frustrated phases and many boundaries, only for the state near the FR$_1$-FR$_2$ boundary is there  a pseudo-transition at a finite temperature. 
This is a good demonstration of the predictive power of the criterion for pseudo-transitions 
formulated earlier \cite{Rojas2020-1,Rojas2020-2}.

The pseudo-critical transitions observed at the phase boundaries offer valuable insights into the interplay between temperature and magnetic field in inducing phase transitions. These findings contribute to a deeper understanding of statistical physics and phase transitions and have implications in various scientific disciplines. Further investigations into this model can open up new avenues for exploring the dynamics of complex systems and phase transitions, enriching the field of condensed matter physics.

\acknowledgments The work was partly supported by the Ministry of Science and Higher Education the Russian Federation (Ural Federal University Program of Development within the Priority-2030 Program), and Brazilian agency CNPq and FAPEMIG.

\appendix

\section{Decoration transformation for $q$-state Potts model\label{sec:Dec-trans}}

The decoration transformation~\cite{Fisher,Syozi,phys-A-09-1,phys-A-09-2} has been
widely used in Ising models and Ising-Heisenberg models. In this appendix,
we apply the decoration transformation mapping to transform the $q$-state
Potts model on a diamond chain to an effective one-dimensional Potts-Zimm-Bragg
model, as considered in reference \cite{Panov}.

To study the thermodynamics of the Hamiltonian \eqref{eq:Ham1}, we
need to obtain the partition function using transfer matrix techniques.
The elements of the transfer matrix are commonly known as Boltzmann
factors.

\begin{alignat}{1}
w(\sigma_{1}^{c},\sigma_{2}^{c})= & {\rm e}^{\frac{\beta h_{1}}{2}\left(\delta_{\sigma_{1}^{c},1}+\delta_{\sigma_{2}^{c},1}\right)}\nonumber \\
 & \sum_{\sigma^{a},\sigma^{b}=1}^{q}\Bigl\{{\rm e}^{\beta J_{ab}\delta_{\sigma^{a},\sigma^{b}}+\beta h_{2}\left(\delta_{\sigma^{a},1}+\delta_{\sigma^{b},1}\right)}\nonumber \\
 & \times{\rm e}^{\beta J_{1}\bigl(\delta_{\sigma_{1}^{c},\sigma^{a}}+\delta_{\sigma_{1}^{c},\sigma^{b}}+\delta_{\sigma^{a},\sigma_{2}^{c}}+\delta_{\sigma^{b},\sigma_{2}^{c}}\bigr)}\Bigr\},\label{eq:wo}
\end{alignat}
the summation in \eqref{eq:wo} can be expressed as
\begin{alignat}{1}
\sum_{\sigma^{a},\sigma^{b}}^{q}\cdots=	& \left(\sum_{\sigma^{a}=1}^{q}{\rm e}^{\beta\left[J_{1}\left(\delta_{\sigma_{1}^{c},\sigma^{a}}+\delta_{\sigma^{a},\sigma_{2}^{c}}\right)+h_{2}\delta_{\sigma^{a},1}\right]}\right)^{2}\nonumber\\
&   +\left(y-1\right)\sum_{\sigma^{a}=1}^{q}{\rm e}^{2\beta\left[J_{1}\left(\delta_{\sigma_{1}^{c},\sigma^{a}}+\delta_{\sigma^{a},\sigma_{2}^{c}}\right)+h_{2}\delta_{\sigma^{a},1}\right]},
\end{alignat}
where we are using the following notation

\begin{equation}
{\rm e}^{\beta J_{ab}\delta_{\sigma_{i}^{a},\sigma_{i}^{b}}}=1+(y-1)\delta_{\sigma_{i}^{a},\sigma_{i}^{b}},
\end{equation}
with $y={\rm e}^{\beta J_{ab}}$.

Thus the Boltzmann factor \eqref{eq:wo}, can be simplified after
some algebraic manipulation

\begin{equation}
w(\sigma_{1}^{c},\sigma_{2}^{c})=\nu_{0}+\nu_{1}\delta_{\sigma_{1}^{c},\sigma_{2}^{c}}+\nu_{2}\delta_{\sigma_{1}^{c},1}\delta_{1,\sigma_{2}^{c}}+\nu_{3}\left(\delta_{\sigma_{1}^{c},1}+\delta_{1,\sigma_{2}^{c}}\right),
\end{equation}
by using the following notation just to express by a simple expression
\begin{eqnarray}
\nu_{0}&= & t_{2}\\
\nu_{1}&= & d_{2}-t_{2}\\
\nu_{2}&= & d_{1}-d_{2}-2\left(t_{1}-t_{2}\right)\\
\nu_{3}&= & t_{1}-t_{2},
\end{eqnarray}
where we have denoted the Boltzmann factors by $w(1,1)=d_{1}$, $w(\mu,\mu)=d_{2}$,
$w(1,\mu)=t_{1}$ and $w(\mu,\mu')=t_{2}$, with $\mu$ and $\mu'$
taking $\{2,3,\dots,q\}$.

On the other hand, based on the Hamiltonian considered in reference
\cite{Panov}, let us write the effective one-dimensional Potts-Zimm-Bragg
model, whose Hamiltonian has the following form

\begin{equation}
\mathsf{H}=-\sum_{i=1}^{N}\left\{ K_{0}+K\delta_{\sigma_{i}^{c},\sigma_{i+1}^{c}}+K_{1}\delta_{\sigma_{i}^{c},1}\delta_{1,\sigma_{i+1}^{c}}+h\delta_{\sigma_{i}^{c},1}\right\} ,\label{eq:H-dec}
\end{equation}
here $K_{0}$, $K$, $K_{1}$ and $h$ must be considered as the effective
parameters.

Therefore, the corresponding Boltzmann factors of the effective model
becomes

\begin{equation}
\mathsf{w}(\sigma_{1}^{c},\sigma_{2}^{c})={\rm e}^{\beta\left\{ K_{0}+K\delta_{\sigma_{1}^{c},\sigma_{2}^{c}}+K_{1}\delta_{\sigma_{1}^{c},1}\delta_{1,\sigma_{2}^{c}}+\frac{h}{2}\left(\delta_{\sigma_{1}^{c},1}+\delta_{1,\sigma_{2}^{c}}\right)\right\} }.
\end{equation}

Using the decoration transformation, we can impose the condition $w(\sigma_{1}^{c},\sigma_{2}^{c})=\mathsf{w}(\sigma_{1}^{c},\sigma_{2}^{c})$.
This results in four non-equivalent algebraic equations that allow
us to determine the four unknown effective parameters of the Hamiltonian
\eqref{eq:H-dec} by solving the system of equations. The solution
are given which result as:

\begin{eqnarray}
K_{0}&= & \frac{1}{\beta}\ln\left[w(\mu,\mu')\right]=\frac{1}{\beta}\ln\left(t_{2}\right)\\
K&= & \frac{1}{\beta}\ln\left[\frac{w(\mu,\mu)}{w(\mu,\mu')}\right]=\frac{1}{\beta}\ln\left(\frac{d_{2}}{t_{2}}\right)\\
K_{1}&= & \frac{1}{\beta}\ln\left[\frac{w(1,\mu)}{w(\mu,\mu')}\right]=\frac{1}{\beta}\ln\left(\frac{t_{1}}{t_{2}}\right)\\
h&= & \frac{2}{\beta}\ln\left[\frac{w(1,1)}{w(\mu,\mu')}\right]=\frac{2}{\beta}\ln\left(\frac{d_{1}}{t_{2}}\right).
\end{eqnarray}
This transformation maps the diamond chain Potts model \eqref{eq:Ham1}
on an effective one-dimensional Potts-Zimm-Bragg model~\citep{Panov}.

\section{Application of Markov chain theory\label{sec:Markov}}

It is possible to construct a mapping of our one-dimensional model
to some Markov chain if we take as the entries of a transition matrix
$P_{\alpha\gamma}$ the conditional probabilities $P(\gamma|\alpha)$
of the state $\gamma=\left|_{\eta_{i+1}}^{\xi_{i+1}}\zeta_{i+1}\right\rangle $
in the $(i+1)$th cell, given that the $i$th cell is in the state
$\alpha=\left|_{\eta_{i}}^{\xi_{i}}\zeta_{i}\right\rangle $. Conditional
probabilities are determined from the Bayes formula $P(\alpha\gamma)=P(\alpha)P(\gamma|\alpha)$,
where, in turn, 
\begin{eqnarray}
P(\alpha) & = & \left\langle \Delta_{i,\alpha}\right\rangle ,\\
P(\alpha\gamma) & = & \left\langle \Delta_{i,\alpha}\Delta_{i+1,\gamma}\right\rangle ,
\end{eqnarray}
and $\Delta_{i,\alpha}$ is a projector on the $\alpha$ state for
the $i$th cell. Using the transfer matrix $V$, built on the states
$\alpha$, we find 
\begin{multline}
\left\langle \Delta_{i,\alpha}\right\rangle =\lim_{N\to\infty}\frac{{\rm Tr}\left(V^{i-1}\Delta_{i,\alpha}V^{N-i+1}\right)}{{\rm Tr}\left(V^{N}\right)}=\\
=\lim_{N\to\infty}\frac{\sum_{k}\left\langle \alpha|\lambda_{k}\right\rangle \lambda_{k}^{N}\left\langle \lambda_{k}|\alpha\right\rangle }{\sum_{k}\lambda_{k}^{N}}=\left\langle \alpha|\lambda_{1}\right\rangle \left\langle \lambda_{1}|\alpha\right\rangle ,\label{eq:Pa}
\end{multline}

\begin{equation}
\left\langle \Delta_{i,\alpha}\Delta_{i+1,\gamma}\right\rangle =\frac{V_{\alpha\gamma}}{\lambda_{1}}\left\langle \gamma|\lambda_{1}\right\rangle \left\langle \lambda_{1}|\alpha\right\rangle .
\end{equation}
Here $\lambda_{1}$ is the maximum eigenvalue of the transfer matrix
$V$. For a positive matrix, the coefficients $v_{\alpha}=\left\langle \alpha|\lambda_{1}\right\rangle $
can be chosen positive, according to Perron's theorem~\cite{gantmakher2000}.
Assuming that 
\begin{equation}
P_{\alpha\gamma}=P(\gamma|\alpha)=\frac{\left\langle \Delta_{i,\alpha}\Delta_{i+1,\gamma}\right\rangle }{\left\langle \Delta_{i,\alpha}\right\rangle },\label{eq:Pab1}
\end{equation}
we obtain 
\begin{equation}
P_{\alpha\gamma}=\frac{V_{\alpha\gamma}v_{\gamma}}{\lambda_{1}\,v_{\alpha}}.\label{eq:Pab2}
\end{equation}
The stochastic properties of the matrix $P_{\alpha\gamma}$ are checked
directly: 
\begin{equation}
\sum_{\gamma}P_{\alpha\gamma}=\frac{1}{\lambda_{1}\,v_{\alpha}}\sum_{\gamma}V_{\alpha\gamma}v_{\gamma}=1.
\end{equation}

Equation~\eqref{eq:Pab2} for constructing a transition matrix is
known in the theory of non-negative matrices~\cite{gantmakher2000},
but the expression~\eqref{eq:Pab1} reveals its physical content
for our model. This allows us to use the results of a very advanced
field of mathematics, the theory of Markov chains. The state of the
system is determined by the stationary probability vector $\mathbf{w}$
of the Markov chain, which can be found from the following equations
\begin{equation}
\sum_{\alpha}w_{\alpha}P_{\alpha\gamma}=w_{\gamma},\quad\sum_{\alpha}w_{\alpha}=1.
\end{equation}
Using~\eqref{eq:Pa}, one can check that $w_{\alpha}=P(\alpha)$,
and if the transfer matrix $V$ is chosen symmetric, then $w_{\alpha}=v_{\alpha}^{2}$.
For the magnetizations, we obtain following expressions 
\begin{equation}
m_{c}=\mathbf{w}\mathbf{m}_{c},\quad m_{ab}=\mathbf{w}\mathbf{m}_{ab},
\end{equation}
where $\alpha$th component of the vector $\mathbf{m}$ equals to
the corresponding magnetization for the state $\alpha$.

The calculation of the transition matrix $P$ involves finding the
maximum eigenvalue $\lambda_{1}$ of the transfer matrix $V$, the
dimension of which for our model is $q^{3}$. However, the dimension
of the matrices can be reduced using the lumpability method for reducing
the size of the state space of Markov chain~\cite{kemeny1976}. We
divide the original set of $m$ states into $M$ groups and find the
lumped transition matrix. Formally, this can be done using matrices
$L_{A\alpha}$ and $R_{\gamma G}$, $\alpha,\gamma=1,\ldots m$, $A,G=1,\ldots M$:
\begin{equation}
P_{AG}=\sum_{\alpha\gamma}L_{A\alpha}P_{\alpha\gamma}R_{\gamma G},\label{eq:PAB}
\end{equation}
where $R_{\gamma G}=1$ if the state $\gamma$ belongs to the group
with the number $G$, and $R_{\gamma G}=0$ otherwise; $L_{A\alpha}=1/\dim(A)$
if the state $\alpha$ belongs to the group with the number $A$ consisting
of $\dim(A)$ elements, and $L_{A\alpha}=0$ otherwise. 

In our problem,
it is natural to divide $m=q^{3}$ states into phases~(\ref{eq:wfFM1}--\ref{eq:wfFRs}),
and also, to complete the set of states, it is necessary to supplement this list with 2 phases, FR$'$ and FR$''$, the states for which have the following form
\begin{equation}
\left|FR'\right\rangle =\prod_{i}\left|\left[_{\mu}^{1}\right]\mu\right\rangle _{i}, \quad 
\left|FR''\right\rangle =\prod_{i}\left|_{\mu_{i}}^{\mu_{i}}1\right\rangle _{i}.
\end{equation}
The energy and entropy at zero temperature have the values
\begin{alignat}{2}
{\rm FR}':\quad & \varepsilon_0 = -\left(2J_{1}+h\right), & \quad & \mathcal{S}_{0} = \ln2 , \\
{\rm FR}'':\quad & \varepsilon_0 = -\left(J_{ab}+h\right), & \quad & \mathcal{S}_{0} = \ln\left(q-1\right) .
\end{alignat}
The states FR$'$ and FR$''$ are not represented in the phase diagrams
of the ground state in Fig.\ref{fig:GS-q_not_3} and \ref{fig:GS-q=3} by their own domains, but
appear as impurities in mixed states at the phase boundaries. 
Thus, for any $q>3$ the matrix $P_{AG}$ will have dimension $M=11$, 
and $M=10$ for $q=3$.

The equilibrium state of the system will correspond to a stationary
probability vector for the lumped Markov chain, 
\begin{equation}
\sum_{A}w_{A}P_{AG}=w_{G},\quad\sum_{A}w_{A}=1,
\end{equation}
and the expressions for magnetizations will not formally change.

The lumpability~\eqref{eq:PAB} can also be applied to the transfer
matrix $V$, if we are only interested in its maximum eigenvalue.
Indeed, the matrix $V$ is non-negative, so according to the Perron-Frobenius
theorem~\cite{gantmakher2000} 
\begin{equation}
\lambda_{1}=\max_{(v\geqslant0)}\min_{1\leqslant\alpha\leqslant m}\frac{\sum_{\beta}V_{\alpha\gamma}v_{\gamma}}{v_{\alpha}}.\label{eq:lamda-max}
\end{equation}
Since the matrix $R$ sums the matrix elements for states from the
group, and the matrix $L$ removes duplicate rows, the value of $\lambda_{1}$
in Eq.~\eqref{eq:lamda-max} will not change after the lumpability.

\begin{figure}[h]
\includegraphics[width=0.35\textwidth]{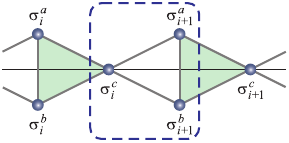} 
\caption{\label{fig:interface} 
An illustration of the interface between $i$th and $(i+1)$th cells of the diamond chain.}
\end{figure}

This computational scheme is greatly simplified for the ground state.
We will count the energy of the system from the energy of the ground
state $E_{0}=N\varepsilon_{0}$. If $T=0$, then for all states with
energy higher than $\varepsilon_{0}$ we get $V_{\alpha\gamma}=0$.
A pair of states $\alpha=\left|_{\eta_{\alpha}}^{\xi_{\alpha}}\zeta_{\alpha}\right\rangle $
and $\gamma=\left|_{\eta_{\gamma}}^{\xi_{\gamma}}\zeta_{\gamma}\right\rangle $
with energy equal to $\varepsilon_{0}$ will be called \emph{allowable}
if the state of the interface $\left|_{\eta_{\gamma}}^{\xi_{\gamma}}\zeta_{\alpha}\right\rangle $
also has energy $\varepsilon_{0}$ (see Fig.\ref{fig:interface}).
For the allowable pair of states, we get $V_{\alpha\gamma}=1$, otherwise
$V_{\alpha\gamma}=0$. For the lumped matrix, the nonzero matrix elements
$V_{AG}$ will be equal to the number of allowable pairs for any state
$\alpha$ from the group $A$ and all states from the group $G$.
As a result, the dimension of a block with nonzero matrix elements
will in most cases be less than $M$.

It can also be shown that for the ground state the entropy per cell
can be calculated as 
\begin{equation}
\mathcal{S}_{0}=\ln\lambda_{1},
\end{equation}
where $\lambda_{1}$ is the maximum eigenvalue of the matrix $V_{\alpha\gamma}$
(or $V_{AG}$) at $T=0$. For a cyclic closed sequence, the probability
of the state $\left(\alpha_{1}\ldots\alpha_{N}\alpha_{1}\right)$
has the following form~\cite{Panov2020}: 
\begin{eqnarray}
P\left(\alpha_{1}\ldots\alpha_{N}\alpha_{1}\right) & = & P(\alpha_{1}|\alpha_{2})P(\alpha_{1}|\alpha_{2})\ldots P(\alpha_{N}|\alpha_{1})\nonumber \\
 & = & \prod_{\alpha\gamma}P(\alpha|\gamma)^{NP(\alpha\gamma)}=p_{0}^{N}.\label{eq:p0-1}
\end{eqnarray}
In the ground state, the equation $\mathcal{S}_{0}=-\ln p_{0}$ is
valid \cite{Panov2022}. Consider the limit of $\ln p_{0}$ at zero
temperature. We have 
\begin{multline}
\ln p_{0}=\sum_{\alpha}P(\alpha\alpha)\ln P(\alpha|\alpha)\;+\\
+\sum_{\alpha<\gamma}P(\alpha\gamma)\ln\left[P(\alpha|\gamma)P(\gamma|\alpha)\right].
\end{multline}
Using the Equations~\eqref{eq:Pab1} and~\eqref{eq:Pab2}, we obtain:
\begin{equation}
P(\alpha|\alpha)=\frac{V_{\alpha\alpha}}{\lambda_{1}},\quad P(\alpha|\gamma)P(\gamma|\alpha)=\frac{V_{\alpha\gamma}V_{\gamma\alpha}}{\lambda_{1}^{2}}.
\end{equation}
The Eq.~\eqref{eq:p0-1} takes the form 
\begin{equation}
\ln p_{0}=\sum_{\alpha\gamma}P(\alpha\gamma)\ln V_{\alpha\gamma}-\ln\lambda_{1}.
\end{equation}
For allowable pairs $P(\alpha\gamma)\neq0$ and $V_{\alpha\gamma}\to1$
at $T\to0$, hence $P(\alpha\gamma)\ln V_{\alpha\gamma}\to0$. For
the other pairs $V_{\alpha\gamma}\to0$, and hence $P(\alpha\gamma)\ln V_{\alpha\gamma}\propto V_{\alpha\gamma}\ln V_{\alpha\gamma}\to0$
at $T\to0$. As a result, $\mathcal{S}\to\ln\lambda_{1}$ at $T\to0$.

\medskip{}

\emph{Example 1.} Consider the ground state at the boundary between
phases FR$_{1}$ and FR$_{4}$. The energies of these phases $\varepsilon_{FR_{1}}=-\left(J_{ab}+2h\right)$
and $\varepsilon_{FR_{4}}=-J_{ab}$ become equal to $\varepsilon_{0}=-J_{ab}$
if $h=0$. The FR$''$ phase has the same energy. These phase states
for the $i$th cell have the form 
\begin{eqnarray}
\left|FR_{1}\right\rangle _{i} & = & \left|_{1}^{1}\mu_{i}\right\rangle ,\\
\left|FR_{4}\right\rangle _{i} & = & \left|_{\nu_{i}}^{\nu_{i}}\mu_{i}\right\rangle ,\\
\left|FR''\right\rangle _{i} & = & \left|_{\mu_{i}}^{\mu_{i}}1\right\rangle ,
\end{eqnarray}
so that 
\begin{equation}
\mathbf{m}_{c}=\begin{pmatrix}0\\0\\1\end{pmatrix},\quad
\mathbf{m}_{ab}=\begin{pmatrix}2\\0\\0\end{pmatrix}.
\end{equation}
The nonzero block of the matrix $V_{AB}$ has the following form:
\begin{equation}
V=\begin{pmatrix}q-1 & (q-2)^{2} & q-2\\
q-1 & (q-2)^{2} & q-2\\
0 & (q-1)(q-2) & q-1
\end{pmatrix}.
\end{equation}
Here is taken into account that the interface state for pairs FR$''$-FR$_{1}$
has the form FM$_{1}$, with energy higher than $\varepsilon_{0}$,
and for pairs, for example, FR$_{1}$-FR$''$ an invalid state FM$_{2}$
may also occur. Finding the maximum eigenvalue of $\lambda_{1}=(q-1)^{2}$,
we get 
\begin{equation}
\mathcal{S}_{0}=2\ln\left(q-1\right),\quad
\mathbf{v} = C\begin{pmatrix}1\\ 1\\ 1\end{pmatrix},
\end{equation}
and calculate the transition matrix 
\begin{equation}
P_{AG}=\frac{V_{AG}}{\lambda_{1}}\frac{v_{G}}{v_{A}}:\;
P=\frac{1}{q-1}\begin{pmatrix}1 & \frac{(q-2)^{2}}{q-1} & \frac{q-2}{q-1}\\
1 & \frac{(q-2)^{2}}{q-1} & \frac{q-2}{q-1}\\
0 & q-2 & 1
\end{pmatrix}.
\end{equation}
Finding the stationary distribution of the lumped Markov chain 
\begin{equation}
P^{T}\mathbf{w}=\mathbf{w}:\quad
\mathbf{w}=\frac{1}{q}\begin{pmatrix}1\\q-2\\1\end{pmatrix},
\end{equation}
we calculate the magnetizations: 
\begin{equation}
m_{c}=\mathbf{w}\mathbf{m}_{c}=\frac{1}{q},\quad m_{ab}=\mathbf{w}\mathbf{m}_{ab}=\frac{2}{q}.
\end{equation}

Note that without taking into account the FR$''$ states, the nonzero magnetization
$m_{c}$ at the phase boundary FR$_{1}$ and FR$_{4}$ looks mysterious,
since for these phases themselves $m_{c}=0$. 
Similarly, the FR$'$ states contribute in a state of FR$_{2}$-FR$_{3}$ phase boundary,
where the energies of these three states are equal. 

\medskip{}

\emph{Example 2.} For the ground state at the boundary between phases
FR$_{2}$ and FR$_{6}$, the energies of these phases $\varepsilon_{FR_{2}}=-2\left(J_{1}+h\right)$
and $\varepsilon_{FR_{6}}=-h$ become equal to $\varepsilon_{0}=2J_{1}$
if $h=-2J_{1}$. The phase FR$_{5}$ has the same energy. These phase
states for the $i$th cell have the form 
\begin{eqnarray}
\left|FR_{2}\right\rangle _{i} & = & \left|\left[_{\mu_{i}}^{1}\right]1\right\rangle ,\\
\left|FR_{6}\right\rangle _{i} & = & \left|\left[_{\nu_{i}}^{1}\right]\mu_{i}\right\rangle ,\\
\left|FR_{5}\right\rangle _{i} & = & \left|_{\nu_{i}}^{\mu_{i}}1\right\rangle ,
\end{eqnarray}
and 
\begin{equation}
\mathbf{m}_{c}=\begin{pmatrix}1\\ 0\\ 1\end{pmatrix},\quad
\mathbf{m}_{ab}=\begin{pmatrix}1\\ 1\\ 0\end{pmatrix}.
\end{equation}
The lumped transfer matrix 
\begin{equation}
V=\begin{pmatrix}2(q-1) & 2(q-1)(q-2) & (q-1)(q-2)\\
2(q-2) & 2(q-2)^{2} & 0\\
2(q-1) & 2(q-1)(q-2) & (q-1)(q-2)
\end{pmatrix},
\end{equation}
has a maximum eigenvalue 
\begin{equation}
\lambda_{1}=\frac{1}{2}\left[3q^{2}-9q+8+\phi_{1}(q)\right],
\end{equation}
where 
\begin{equation}
\phi_{1}(q)=\sqrt{q^{4}+2q^{3}-15q^{2}+16q}.
\end{equation}
We write the corresponding eigenvector 
\begin{equation}
\mathbf{v}=C\begin{pmatrix}1\\
\dfrac{(q-2)\left(q^{2}-3q+4+\phi_{1}(q)\right)}{(q-1)\left(3q^{2}-9q+8+\phi_{1}(q)\right)}\\
1 \end{pmatrix},
\end{equation}
and find the transition matrix: 
\begin{equation}
P=\begin{pmatrix}\frac{2(q-1)}{\lambda_{1}} & \frac{2(q-2)^{2}\left(\lambda_{1}-q^{2}+3q-2\right)}{\lambda_{1}^{2}} & \frac{(q-2)(q-1)}{\lambda_{1}}\\
\frac{2(q-1)}{\lambda_{1}-q^{2}+3q-2} & \frac{2(q-2)^{2}}{\lambda_{1}} & 0\\
\frac{2(q-1)}{\lambda_{1}} & \frac{2(q-2)^{2}\left(\lambda_{1}-q^{2}+3q-2\right)}{\lambda_{1}^{2}} & \frac{(q-2)(q-1)}{\lambda_{1}}\\
\end{pmatrix}.
\end{equation}
The stationary probabilities can be reduced to the form 
\begin{equation}
\mathbf{w}=\frac{1}{2\phi_{1}(q)}\begin{pmatrix}4(q-1)\\
\phi_{1}(q)+q^{2}-7q+8\\
\phi_{1}(q)-q^{2}+3q-4
\end{pmatrix},
\end{equation}
that allow us to find the magnetizations: 
\begin{equation}
m_{c}=\frac{\phi_{1}(q)-q^{2}+7q-8}{2\phi_{1}(q)},
\end{equation}

\begin{equation}
m_{ab}=\frac{\phi_{1}(q)+q^{2}-3q+4}{2\phi_{1}(q)}.
\end{equation}

In this way, it is possible to obtain all the values given in the
Table~\ref{tab:GS-En-Entr-M}.

\medskip

A special situation occurs at $q=3$, when the ground state energy has the value $\varepsilon_{0}=-h$. 
This value has the energy of the FR$_{5}$ and FR$_{6}$ phases. 
At $q>3$, the entropy of the FR$_{6}$ phase is greater than that of the FR$_{5}$ phase, so when $T>0$, the free energy for the FR$_{6}$ phase is less than for the FR$_{5}$ phase, and in the limit $T\to0$, the main state is the FR$_{6}$ phase. 
If $q=3$, the entropy of FR$_{5}$ and FR$_{6}$ phases is equal to $\mathcal{S}_{0}=\ln2$. 
The nature of the ground state in this case can be investigated using the Markov chain method proposed above. 

At a sufficiently low temperature, the state of the system is formed by phases whose energies are closest to the energy of the ground state. 
In the parameter range $h>0$, $J_{1}<-h/2$, and $J_{ab}<-h$ (see Fig.\ref{fig:GS-q=3}a), it is natural to take into account in addition to FR$_{5}$ and FR$_{6}$ phases also neighboring phases FR$_{1}$ and FR$_{2}$.
The transfer matrix with enties $V_{AB}$, where $A,B={\rm FR}_{1},\,{\rm FR}_{2},\,{\rm FR}_{5},\,{\rm FR}_{6}$, has the form
\begin{equation}
	V = 2z
\begin{pmatrix}
 yz & \sqrt{xy}z\left(1+x\right) & x \sqrt{yz} & \sqrt{yz}\left(1+x\right) \\
 x^{5/2} \sqrt{y} z & 2 x^2 z & \sqrt{xz} & 2 x^{3/2} \sqrt{z} \\
 x^2 \sqrt{yz} & 2 x^{3/2} \sqrt{z} & 1 & 2 x \\
 \sqrt{yz} & \sqrt{xz}\left(1+x\right) & x & 1+x 
\end{pmatrix}.
\end{equation}
Here $x={\rm e}^{\beta J_{1}}$, $y={\rm e}^{\beta J_{ab}}$, $z={\rm e}^{\beta h}$, and it is taken into account that $q=3$. 
Explicit expressions for the maximum eigenvalue $\lambda_1$ and its eigenvector $\mathbf{v}$ have a rather cumbersome form, however, for the parameters under consideration and $\beta\gg1$, their approximate expressions can be used: 
\begin{equation}
	\lambda_1 = 2z \left(1+u\right), \quad 
	\mathbf{v} 
	= C \begin{pmatrix} \sqrt{yz} \\ 2x^{3/2}\sqrt{z}/u \\ 2x/u \\ 1 \end{pmatrix} , 
\end{equation}
where 
\begin{equation}
	u = \frac{1}{2} \left( yz + \sqrt{y^2 z^2 + 8 x^2 z } \right) . 
\end{equation}
Using these expressions in Eq.~\eqref{eq:Pab2} and leaving only the leading terms for $\beta\gg1$ in the entries of the matrix $V$, we obtain the transition matrix:
\begin{equation}
	P = \frac{1}{1+u}
\begin{pmatrix}
 yz & \dfrac{2 x^2z}{u} & \dfrac{2 x^2}{u} & 1 \\
 \frac{1}{2} uxyz & 2 x^2 z & 1 & u \\
 \frac{1}{2} uxyz & 2 x^2 z & 1 & u \\
 yz & \dfrac{2 x^2z}{u} & \dfrac{2 x^2}{u} & 1 
\end{pmatrix}.
\end{equation}
In the parameter domain under consideration, the stochastic properties of this matrix are hold quite accurate at $\beta\gg 1$. 

The qualitative difference of Markov chains generated by $P$ at low temperature depends on the asymptotic behavior of the parameter $u$: 
\begin{equation}
	u \xrightarrow[\beta\gg1]{} 
	\left\{
	\begin{array}{ll}
		yz , & 2 J_{ab} + h > 2 J_{1} , \\
		2 yz = 2 x \sqrt{z} , & 2 J_{ab} + h = 2 J_{1} , \\
		x\sqrt{2z} , & 2 J_{ab} + h < 2 J_{1} . \\
	\end{array}
	\right.
\end{equation}

\begin{figure}
\includegraphics[width=0.9\linewidth]{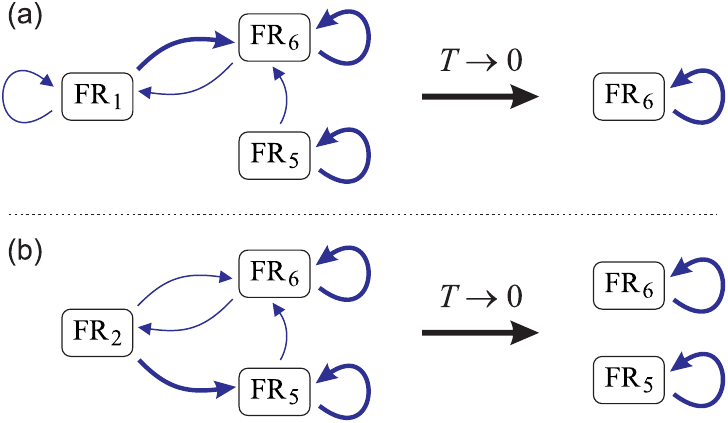} 
\caption{\label{fig:Graphs} 
Transition graphs of Markov chains at $q=3$, $h>0$, $2J_{1}<-h$, $J_{ab}<-h$ 
and low temperature for 
(a) $2J_{ab}+h>2J_{1}$, (b) $2J_{ab}+h<2J_{1}$. }
\end{figure}

Consider the case of $2J_{ab}+h>2J_{1}$. 
Under this condition, the following inequalities will be true: 
$x^2 \ll x^2z \ll yz$, $uxyz \ll yz$. 
For clarity, we leave in the matrix $P$ only entries of order 1 and the first order of smallness, and replace matrix elements of higher orders of smallness with zeros. 
As a result, the transition matrix takes the form 
\begin{equation}
	P =
\begin{pmatrix}
 a_1 & 0 & 0 & 1-a_1 \\
 0 & 0 & 1-a_1 & a_1 \\
 0 & 0 & 1-a_1 & a_1 \\
 a_1 & 0 & 0 & 1-a_1 
\end{pmatrix}, \quad a_1 = yz . 
\end{equation}
Transition graph of this Markov chain is shown in Fig.\ref{fig:Graphs}a. 
The state FR$_{2}$ in this case is transient and is omitted for simplicity. 
Thin and thick lines correspond to the transition probabilities $a_1$ and $1-a_1$. 
The stationary state contains an exponentially small admixture of the FR$_{1}$ phase 
and in the limit of $T\to0$ becomes a pure FR$_{6}$ phase:
\begin{equation}
	\mathbf{w} = \begin{pmatrix} a_1 \\ 0 \\ 0 \\ 1-a_1 \end{pmatrix} \; 
	\xrightarrow[T\to0]{} \; 
	\begin{pmatrix} 0 \\ 0 \\ 0 \\ 1 \end{pmatrix} . 
\end{equation}

If $2 J_{ab} + h > 2 J_{1}$, when $u \approx x\sqrt{2z}$, the estimates hold 
$uxyz \ll yz \ll u$, $x^2 \ll u$, and $2x^2z/u  \approx u$. 
Replacing exponentially small entries with zeros, we get the transition matrix 
\begin{equation}
	P =
\begin{pmatrix}
 0 & a_2 & 0 & 1-a_2 \\
 0 & 0 & 1-a_2 & a_2 \\
 0 & 0 & 1-a_2 & a_2 \\
 0 & a_2 & 0 & 1-a_2 
\end{pmatrix}, \quad a_2 = x\sqrt{2z} . 
\end{equation}
The corresponding graph without the transient state FR$_{1}$ is shown in Fig.\ref{fig:Graphs}b. 
Thin and thick lines correspond to the transition probabilities $a_2$ and $1-a_2$. 
The stationary state in this case contains an exponentially small admixture of the FR$_{2}$ phase and at $T\to0$ transforms into a mixture of independent phases FR$_{5}$ and FR$_{6}$ having equal fractions: 
\begin{equation}
	\mathbf{w} = \frac{1}{2} \begin{pmatrix} 0 \\ a_2 \\ 1-a_2 \\ 1 \end{pmatrix} \; 
	\xrightarrow[T\to0]{} \; 
	\frac{1}{2} \begin{pmatrix} 0 \\ 0 \\ 1 \\ 1 \end{pmatrix} . 
\end{equation}
It is this state that is designated as ${\rm FR}_{5}+{\rm FR}_{6}$ in Fig.\ref{fig:GS-q=3}a. 

On the boundary $2J_{ab}+h=2J_{1}$, similar considerations give a transition matrix
\begin{equation}
	P =
\begin{pmatrix}
 a_3 & a_3 & 0 & 1-2a_3 \\
 0 & 0 & 1-2a_3 & a_3 \\
 0 & 0 & 1-2a_3 & a_3 \\
 a_3 & a_3 & 0 & 1-2a_3 
\end{pmatrix}, \; a_3 = x\sqrt{z} = yz . 
\end{equation}
The stationary state in this case transforms at $T\to 0$ into a mixture of independent phases FR$_{5}$ and FR$_{6}$ with a ratio of fractions of $1/2$:
\begin{equation}
	\mathbf{w} = \frac{1}{3} \begin{pmatrix} 2a_3 \\ 2a_3 \\ 1-2a_3 \\ 2-2a_3 \end{pmatrix} \; 
	\xrightarrow[T\to0]{} \; 
	\frac{1}{3} \begin{pmatrix} 0 \\ 0 \\ 1 \\ 2 \end{pmatrix} . 
\end{equation}

At $h<0$, similar results for the composition of the phases of the ground state, 
shown in Fig.\ref{fig:GS-q=3}b, can be obtained taking into account the mixing of the phases FR$_{5}$ and FR$_{6}$ states of neighboring phases FR$_{3}$ and FR$_{4}$. 
A special composition of the ground state also occurs at $h=0$ in the region of phases FR$_{6}$ and ${\rm FR}_{5}+{\rm FR}_{6}$ in Figures~\ref{fig:GS-q=3}d and~\ref{fig:GS-q=3}f. 
The stationary state at $T=0$ on the line $h=0$ is a mixture of FR$_{5}$ and FR$_{6}$ with a ratio of fractions of $1/2$. 


%

\end{document}